\documentclass[usenatbib,usegraphicx,usedcolumn]{mn2e}
\usepackage{times,amssymb,amsmath}
\usepackage{relsize}

\usepackage[backref,breaklinks,colorlinks,citecolor=blue]{hyperref}
\usepackage[usenames,dvipsnames]{color}
\title[Moving Mesh Lloyd Regularization]{Reducing noise in moving-grid codes with strongly-centroidal Lloyd mesh regularization}
\author[P. Mocz et. al.]{Philip Mocz$^{1}$\thanks{E-mail: pmocz@cfa.harvard.edu (PM)}, 
Mark Vogelsberger$^{2}$,
R\"udiger Pakmor$^{3}$,
Shy Genel$^{4}$\thanks{Hubble Fellow},
Volker Springel$^{3,5}$, \newauthor
Lars Hernquist$^{1}$ \\
$^{1}$Harvard-Smithsonian Center for Astrophysics, 60 Garden Street, Cambridge, MA 02138, USA \\
$^{2}$Department of Physics, Kavli Institute for Astrophysics and Space Research, Massachusetts Institute of Technology, Cambridge, MA 02139, USA\\
$^{3}$Heidelberger Institut f\"ur Theoretische Studien, Schloss-Wolfsbrunnenweg 35, 69118 Heidelberg, Germany\\
$^{4}$Department of Astronomy, Columbia University, 550 West 120th Street, New York, NY 10027, US\\
$^{5}$Zentrum für Astronomie der Universität Heidelberg, Astronomisches
Recheninstitut, Mönchhofstr. 12-14, 69120 Heidelberg, Germany\\
}

\begin{document}

\date{submitted to MNRAS, Mar 2015} 

\pagerange{\pageref{firstpage}--\pageref{lastpage}} \pubyear{2013}

\maketitle

\label{firstpage}

\begin{abstract}
A method for improving the accuracy of hydrodynamical codes that use a moving Voronoi mesh is described. Our scheme is based on a new regularization scheme that constrains the mesh to be centroidal to high precision while still allowing the cells to move approximately with the local fluid velocity, thereby retaining the quasi-Lagrangian nature of the approach. Our regularization technique significantly reduces mesh noise that is attributed to changes in mesh topology and deviations from mesh regularity. We demonstrate the advantages of our method on various test problems, and note in particular improvements obtained in handling shear instabilities, mixing, and in angular momentum conservation. Calculations of adiabatic jets in which shear excites Kelvin Helmholtz instability show reduction of mesh noise and entropy generation. In contrast, simulations of the collapse and formation of an isolated disc galaxy are nearly unaffected, showing that numerical errors due to the choice of regularization do not impact the outcome in this case.
\end{abstract}


\section{Introduction}\label{sec:intro}

New computational methods for fluid dynamics that employ a moving
Voronoi mesh approach have been developed in the past several years to
simulate astrophysical and cosmological systems.  Codes based on this
method include {\sc AREPO} \citep{2010MNRAS.401..791S}, which has
recently been used to run the 12 billion resolution element
state-of-the-art ``Illustris'' cosmological simulation
\citep{2014Natur.509..177V,2014MNRAS.444.1518V,2014MNRAS.445..175G},
      {\sc TESS} \citep{2011ApJS..197...15D}, adapted for relativistic
      hydrodynamics, the moving mesh algorithm presented in
      \cite{2012ApJ...758..103G} for the simulation of magnetically
      levitating accretion disks around supermassive black holes, and
      the {\sc RICH} code \citep{2015ApJS..216...14S}. The moving
      Voronoi framework has also been extended to finite-element
      techniques \citep{2014MNRAS.437..397M} and constrained transport
      approaches for magnetohydrodynamics (MHD) to strictly maintain
      the divergence-free nature of the magnetic field
      \citep{2014MNRAS.442...43M}. Related mesh-free methods have also
      been developed by \cite{2014arXiv1409.7395H}, which use Riemann
      solvers acting over volume ``overlaps'', and have similar
      advantages to the moving mesh approach, and have particularly
      good angular momentum conservation but at the cost of enhanced
      noise.

Moving mesh codes have found success due to the numerical advantages
they provide because of their quasi-Lagrangian nature,
Galilean-invariant formulation (for non-relativistic fluids), limited
advection errors, preservation of contact discontinuities, continuous
spatial adaptability, and ability to accurately resolve instabilities.
Notably, the moving mesh method provides advantages over previous
smoothed particle hydrodynamics (SPH) approaches, which can suppress
entropy generation by mixing, underestimate vorticity generation in
curved shocks, prevent efficient gas stripping from infalling
substructures \citep{2012MNRAS.424.2999S}, and have relatively poor
convergence properties (e.g. \citealt{2015ApJ...800....6Z}); and also
over adaptive mesh refinement (AMR) codes, which can have large
advection errors due to numerical diffusion in the presence of large
supersonic bulk velocities.

However, moving Voronoi mesh codes can be affected by noise on small
spatial scales due to the mesh motion
\citep{2012MNRAS.423.2558B,2014arXiv1409.7395H}. As the mesh evolves,
the orientation and sizes of the faces and the topology of the
connections between cells change, introducing small truncation errors
in the solution, which can lead to artificial additional power on the
smallest scales \citep{2012MNRAS.423.2558B} and numerical secondary
instabilities in shear flows. The errors are largest in the presence
of shear, when the mesh cells move past each other and change face
boundaries rapidly, because the standard moving mesh formulations
assume that the face orientations do not evolve significantly in one
timestep or may not adequately capture rapid changes. If the mesh
generating points strictly move with the fluid velocity, the mesh
noise can be quite large because the Voronoi cells can deform into
highly irregular shapes.

Mesh regularization, i.e., applying a small correction to the mesh
generating points of the cells in a way that keeps the cells
``well-behaved'' (round and centroidal) provides a handle on this
error. The mesh generating points travel with the fluid velocity plus
some small correction to keep the mesh approximately regular and
centroidal. The approach taken by \cite{2010MNRAS.401..791S} adds a
small correction in the direction of the center-of-mass at the
beginning of the time step (or sub time step in the case of
higher-order time integration). \cite{2012MNRAS.425.3024V} modify this
approach for cosmological simulations by considering the maximum
opening angle of a face at the mesh generating point and the density
gradient direction to set the correction. The correction term scales
as the sound-speed inside the cell, which ensures a timestep
independent formulation. We will refer to this class of regularization
methods as sound-speed regularization (SSR). 

Despite keeping the cells approximately regular, these techniques do
not eliminate mesh noise entirely because there are several small
sources of error present in the formulation.  First, the motion of the
mesh generating points (and hence the change in the orientation of the
faces) is not smooth.  The direction of the correction only takes into
account the geometry at the beginning of the time step.  After the
mesh advances to the next timestep, a small offset between the mesh
generating point and the center-of-mass remains, which may point in an
uncorrelated direction from that at the beginning of the time
step. The cell then obtains an small kick in that uncorrelated
direction, adding a small level of random noise to the smooth mesh
deformation.  Second, a small offset exists in the center-of-mass and
mesh generating point.  This causes the cells to have some spin about
their center of mass, and also introduces errors in the second-order
estimates of cell gradients \citep{2010MNRAS.401..791S}, which assumes
the two points coincide.  This later issue has recently been resolved
in \cite{2015arXiv150300562P}, where an improved least-squares-fit
gradient estimator has been developed that can achieve second-order
accuracy for smooth flows.

Recently, \cite{2014arXiv1407.7300D} proposed a smoothing of the
velocities of the mesh generating points to alleviate the problem of
mesh noise. Their scheme exhibits a reduction in: grid noise,
numerical secondary instabilities, and artificial power in the
power-spectrum of the solution field on the smallest spatial scales
(although not at the level of fixed grid codes). The smoothing method
overcomes the issue of the uncorrelated randomly directed corrections
that violate a fully smooth mesh motion.  However, setting the cell
vertex velocities to the fluid velocity and smoothing by the
neighbours does not enforce a strongly centroidal mesh (only
approximately centroidal; as do the other regularization methods in
the literature); deviations from a centroidal mesh introduce errors in
e.g. Green-Gauss gradient estimates. Moreover, in the presence of
shear flows, the cell velocities in this approach will be smoothed
out, reducing the code to a static mesh code, and thus losing the
desired properties of a moving mesh code such as advecting the
solution at high precision.

We have developed a new regularization scheme to address the
smoothness and strong-centroidality issues, which is presented here.
We refer to the method as strongly-centroidal Lloyd regularization
(LR).  Our approach evolves the mesh in a smooth, centroidal manner.
Unlike the previous techniques, it accounts for where the
center-of-mass of a cell travels to at the \textit{end} of a time step
in order to ensure the centroidal property, making it in essence a
forward-predicting, iterative Lloyd's algorithm.  The concept is
straightforward.  The velocities of mesh generating points are
initialized to the values of the local fluid velocity, and a few
iterations of corrections are applied to modify this velocity so that
the cell moves to the location of its center-of-mass at the end of the
time step.  This results in a smoothly evolving centroidal Voronoi mesh
that moves approximately with the fluid velocity.  The scheme may
optionally be modified to a weighted centroidal scheme (weighted by
the fluid density) for simulations with collapsing structure to gain
automatic increased resolution in regions of high density.  We
demonstrate significant reduction of mesh noise with the new
regularization scheme.

Developing reliable numerical codes that offer general adaptability to
a huge spatial and temporal dynamic range is an ongoing
challenge. Recently \cite{2015arXiv150300562P} have improved the
accuracy and angular momentum conservation for moving-mesh codes by
developing new time integration scheme and spatial gradient
estimates in the {\sc AREPO} code, significantly improving the
precision of the code in smooth test problems.  In that work, the
MUSCL-Hancock (MH) scheme \citep{tororiemann} is replaced by a
second-order, time symmetric Runge-Kutta (RK) integrator via Heun's
method.  Additionally, the Voronoi-improved Green-Gauss (GG) gradient
estimates are replaced by a least-squares gradient estimator
(LSF).  Our regularization complements these recent improvements for
smooth flows by improving the accuracy of moving mesh algorithms when
shear is present.

We describe the method and its computationally efficient
implementation in Section~\ref{sec:method}. Several numerical tests are
performed to show the advantages of our approach, the results of which
are presented in Section~\ref{sec:tests}. We offer concluding remarks in
Section~\ref{sec:conclusion}.

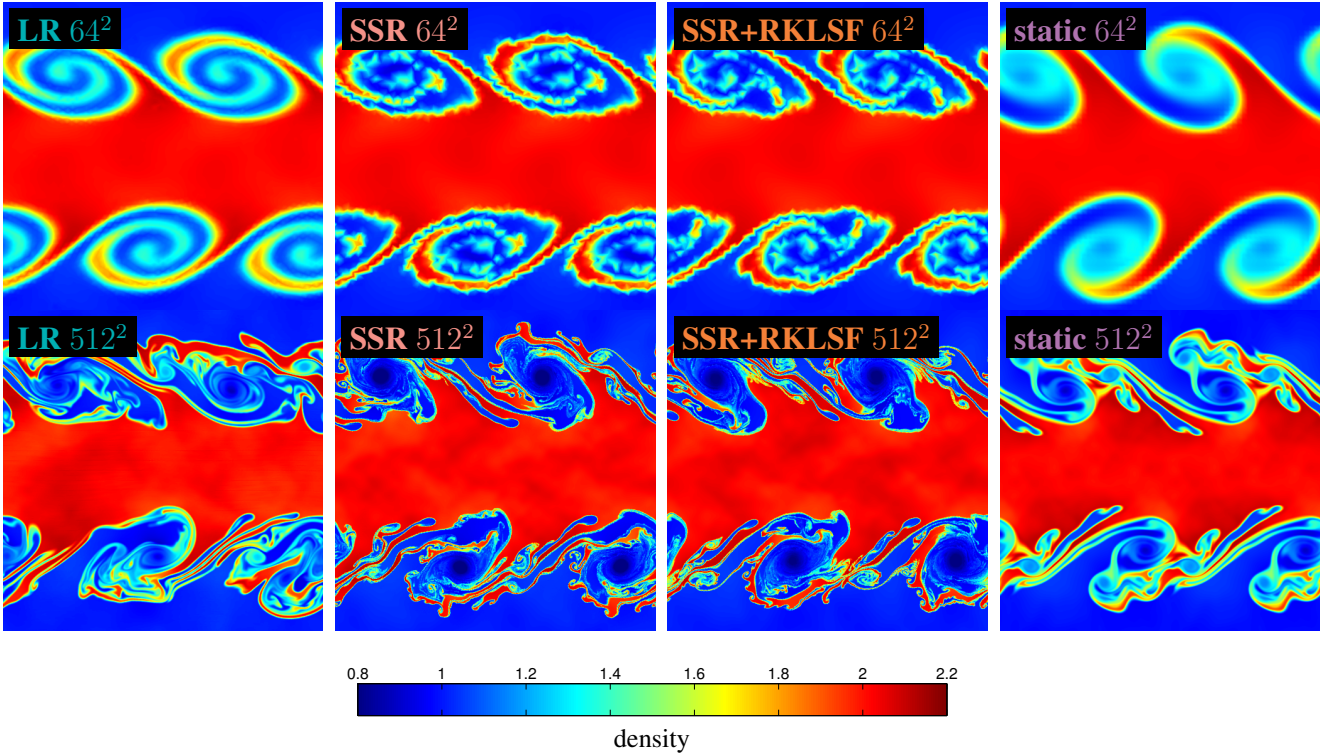
\begin{figure*}
\centering
\begin{tabular}{cccc}
\begin{picture}(0.48,0.48)
\put(0,0) {\includegraphics[width=0.24\textwidth]{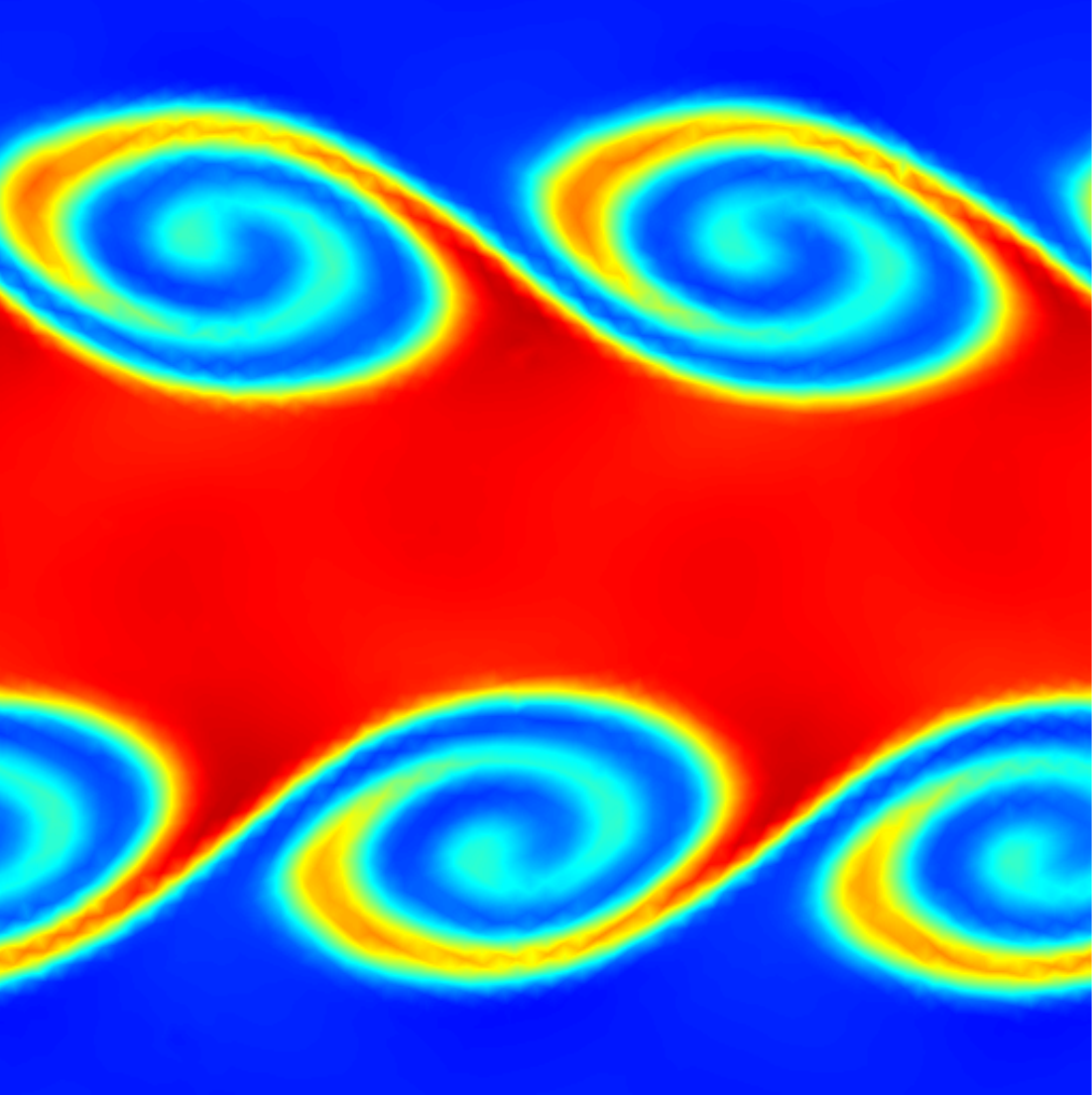} }
\put (0.01,0.45) {\colorbox{black}{\Large\bf\color{TealBlue} LR $64^2$}} 
\end{picture} &
\begin{picture}(0.48,0.48)
\put(0,0) {\includegraphics[width=0.24\textwidth]{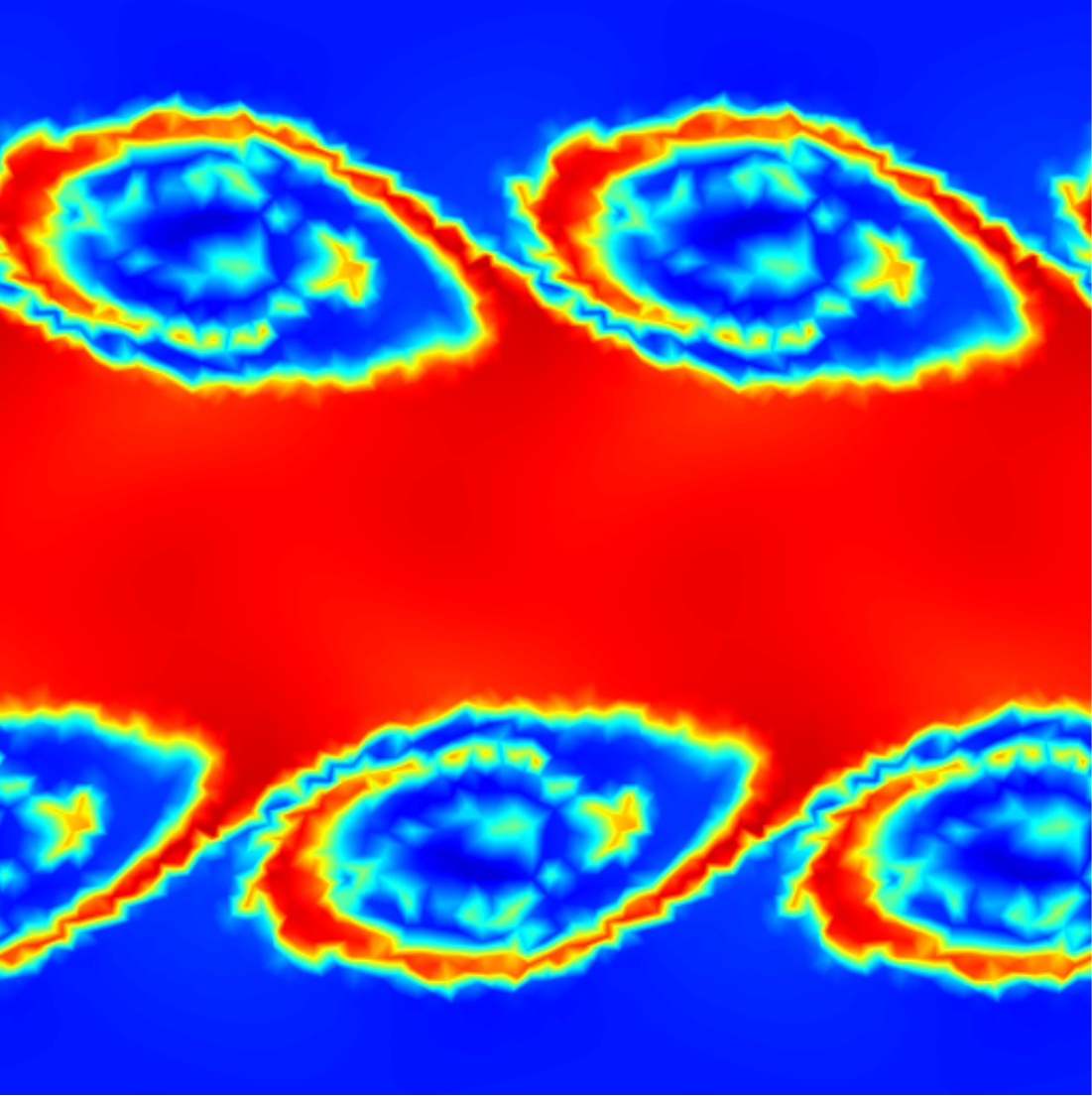} }
\put (0.01,0.45) {\colorbox{black}{\Large\bf\color{Salmon} SSR $64^2$}} 
\end{picture} & 
\begin{picture}(0.48,0.48)
\put(0,0) {\includegraphics[width=0.24\textwidth]{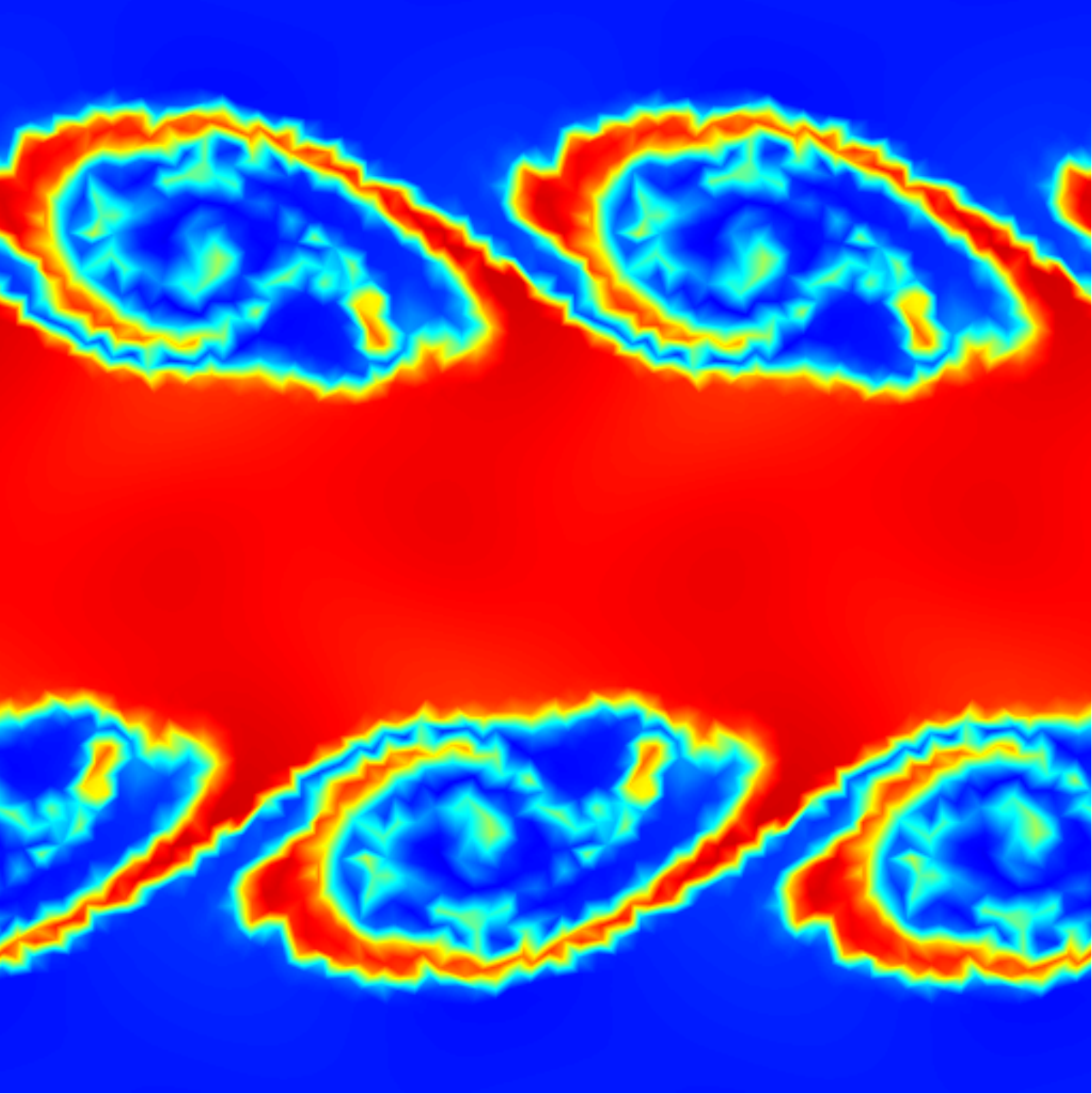} }
\put (0.01,0.45) {\colorbox{black}{\Large\bf\color{Orange} SSR+RKLSF $64^2$}} 
\end{picture} &
\begin{picture}(0.48,0.48)
\put(0,0) {\includegraphics[width=0.24\textwidth]{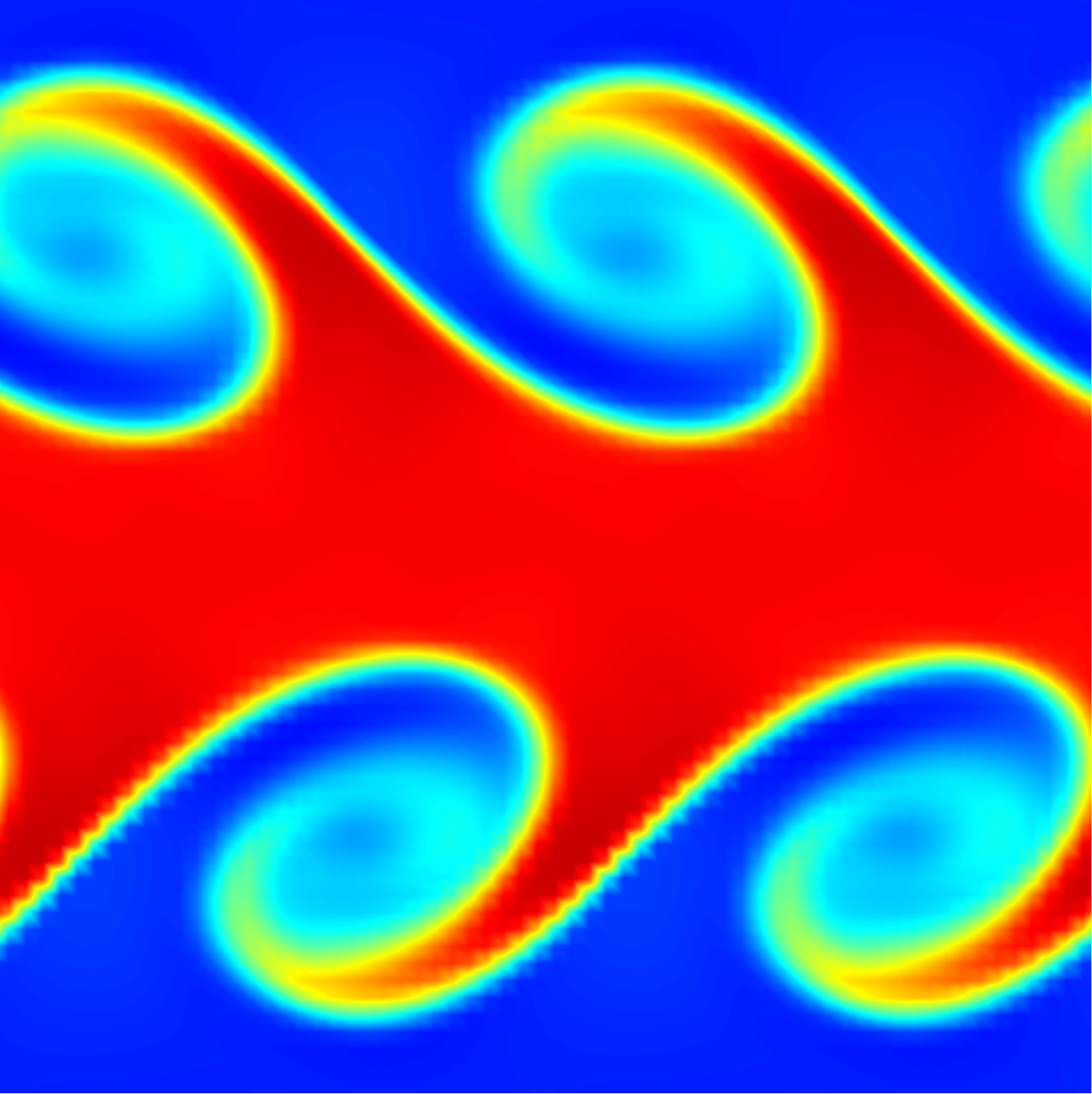} }
\put (0.01,0.45) {\colorbox{black}{\Large\bf\color{Orchid} static $64^2$}} 
\end{picture} \\
\begin{picture}(0.48,0.48)
\put(0,0) {\includegraphics[width=0.24\textwidth]{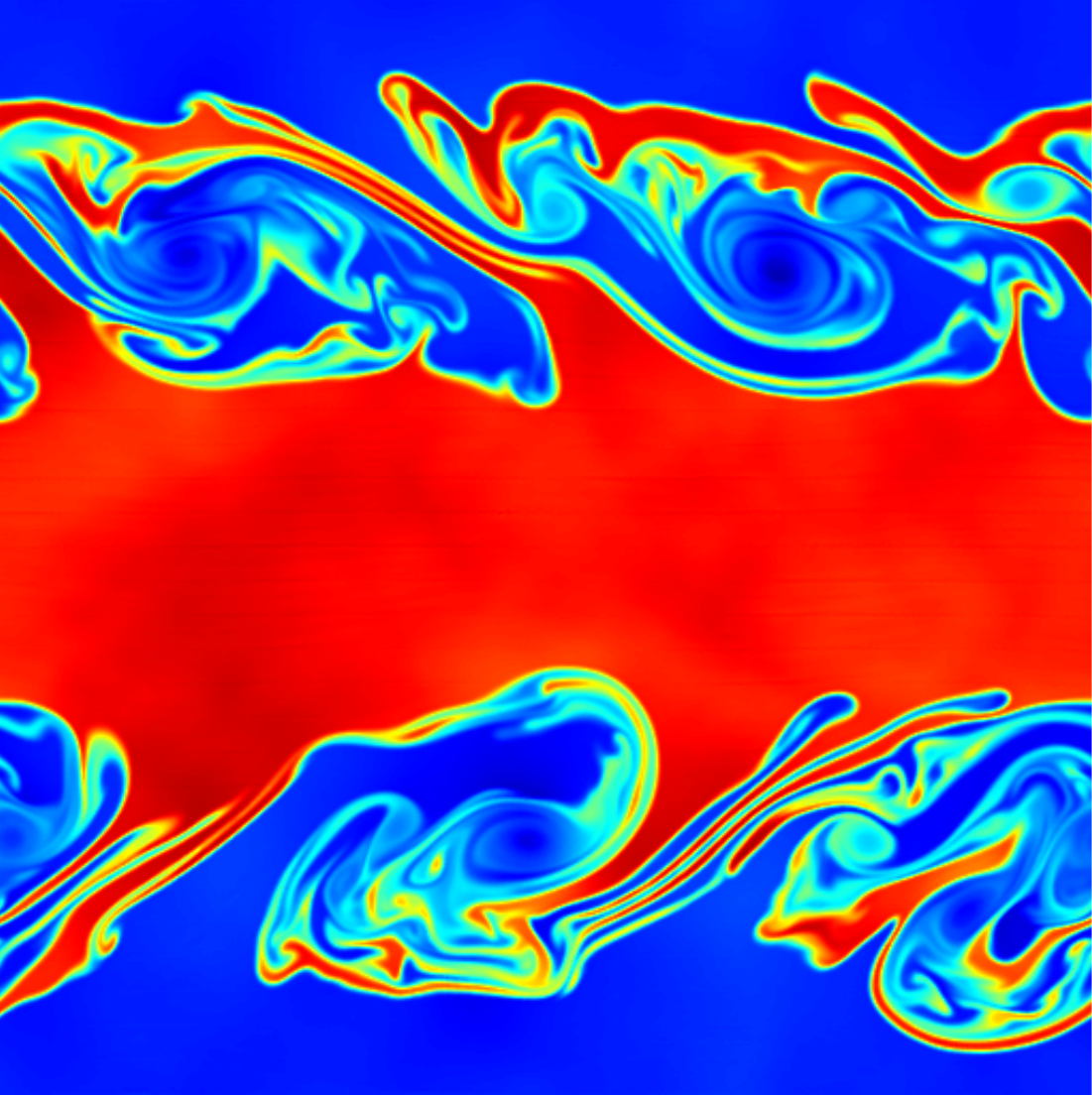} }
\put (0.01,0.45) {\colorbox{black}{\Large\bf\color{TealBlue} LR $512^2$}} 
\end{picture} &
\begin{picture}(0.48,0.48)
\put(0,0) {\includegraphics[width=0.24\textwidth]{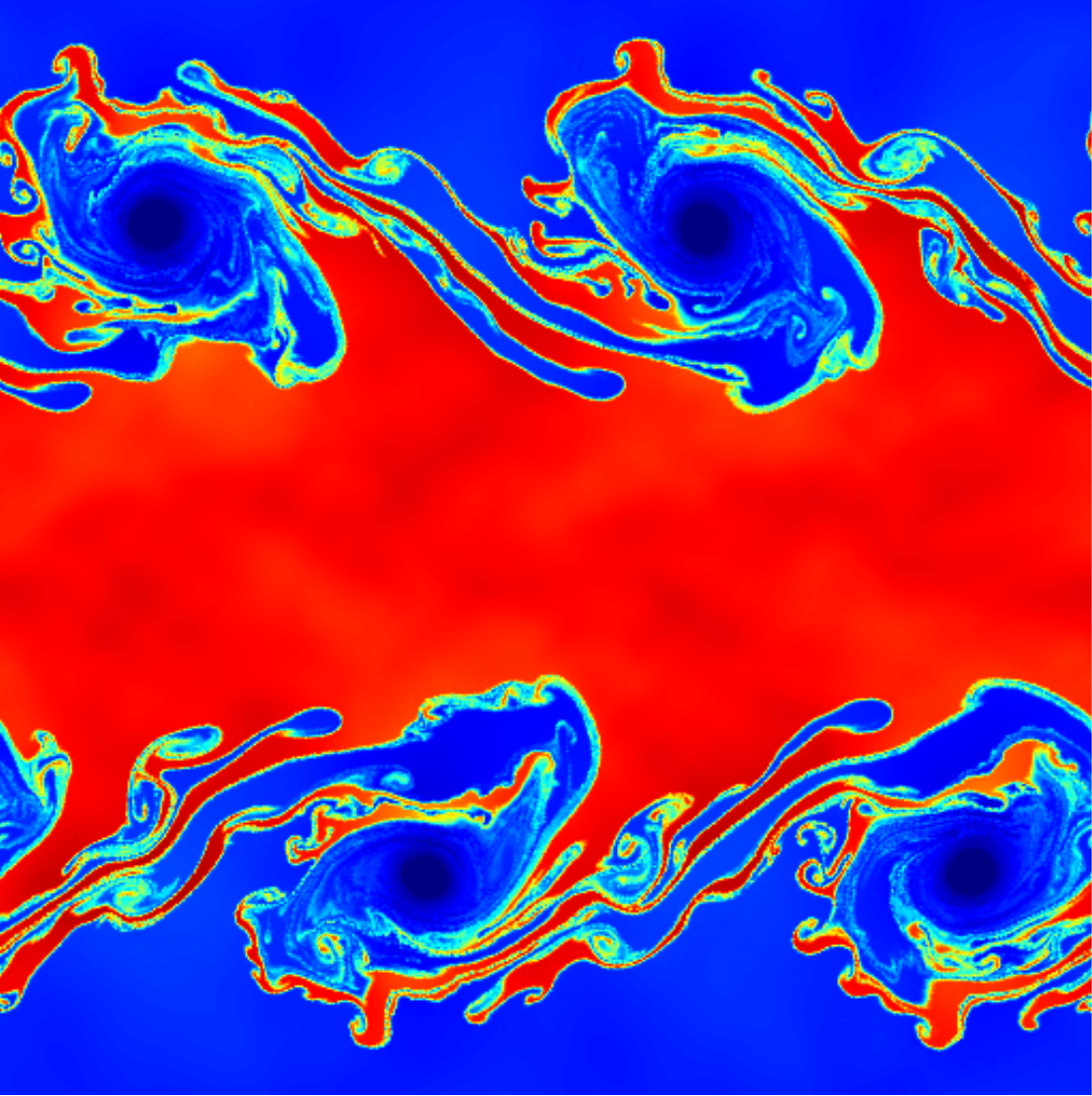} }
\put (0.01,0.45) {\colorbox{black}{\Large\bf\color{Salmon} SSR $512^2$}}
\end{picture} &
\begin{picture}(0.48,0.48)
\put(0,0) {\includegraphics[width=0.24\textwidth]{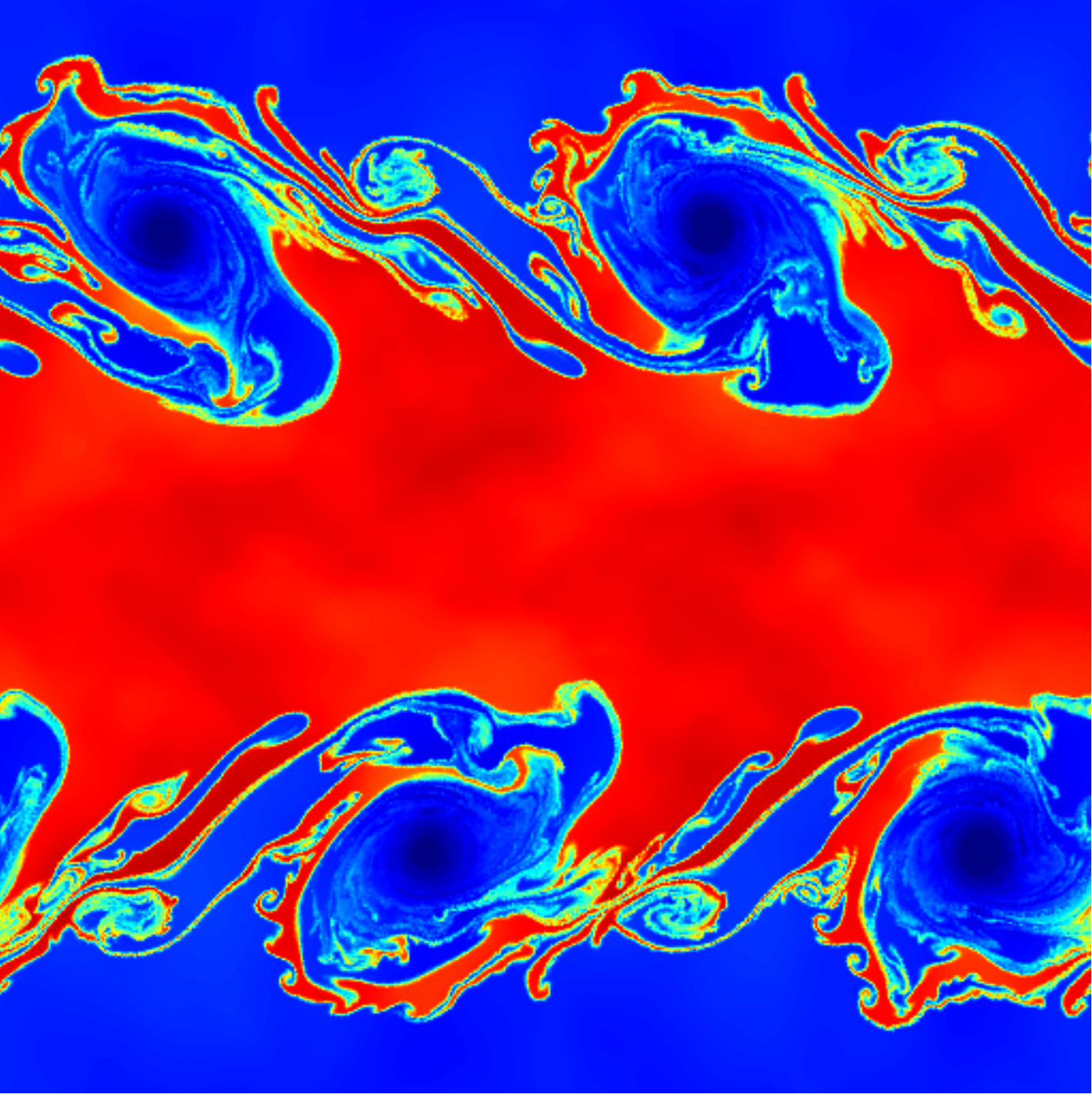} }
\put (0.01,0.45) {\colorbox{black}{\Large\bf\color{Orange} SSR+RKLSF $512^2$}} 
\end{picture} &
\begin{picture}(0.48,0.48)
\put(0,0) {\includegraphics[width=0.24\textwidth]{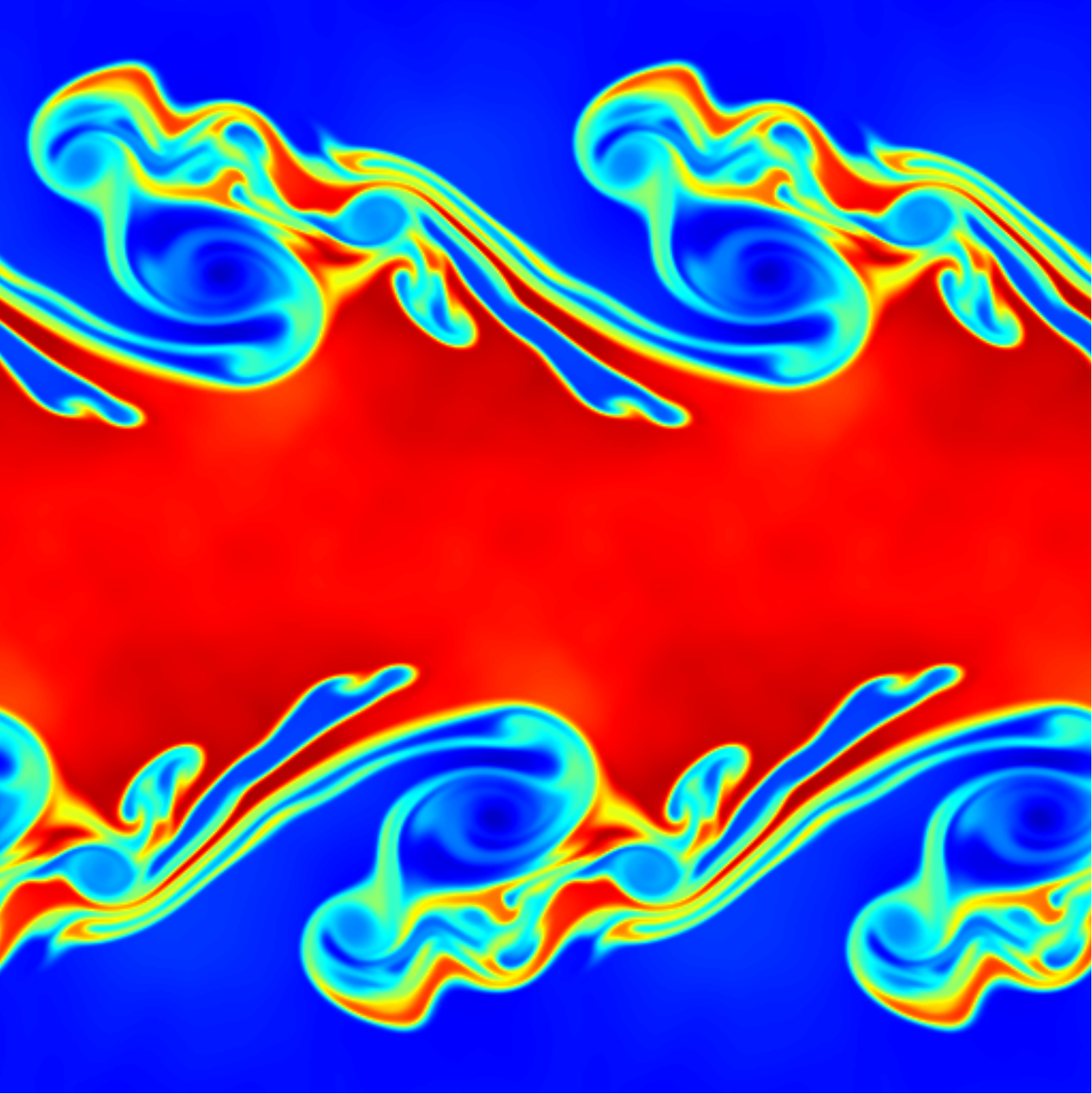} }
\put (0.01,0.45) {\colorbox{black}{\Large\bf\color{Orchid} static $512^2$}} 
\end{picture} \\
\multicolumn{4}{c}{\includegraphics[width=0.48\textwidth]{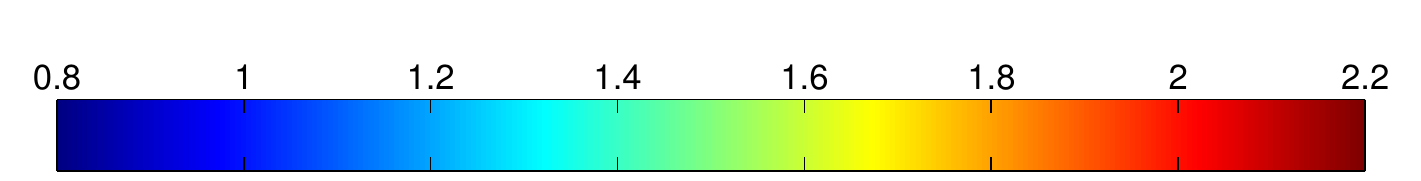}} \\
\multicolumn{4}{c}{{\large density}}
\end{tabular}
\caption{Comparison of KHI with the new (LR) and previous (SSR) regularization schemes at low and high resolution at $t=2.0$. LR eliminates secondary numerical instabilities, which are present in the simulations that use SSR, even with the recent RK and LSF improvements. For comparison, runs on a static grid are also shown, which also do not have secondary numerical instabilities.}
\label{fig:KH}
\end{figure*}

\section{Strongly-centroial Lloyd Regularization}\label{sec:method}

The regularization scheme involves initially setting the velocity of
the mesh generating point $\mathbf{w}_i$ of each cell $i$ equal to
the local fluid velocity $\mathbf{v}_i$:
\begin{equation}
\mathbf{w}_{i,0} =\mathbf{v}_i
\label{eqn:init}
\end{equation}
Then, in each successive iteration $J$, the predicted locations of the
centers-of-mass of the cells $\mathbf{\tilde{c}}_{i,J}$ are computed at
the end of the timestep $\Delta t$, and the mesh generating point
velocities are updated to:
\begin{equation}
\mathbf{w}_{i,J} = \left( \mathbf{\tilde{c}}_{i,J} - \mathbf{r}_i \right)/\Delta t
\end{equation}
where $\mathbf{r}_i$ is the location of cell $i$ at the beginning of the timestep.

This iteration is essentially a Lloyd's algorithm for converging an
arbitrary Voronoi diagram to a centroidal one. We take $5$ iterations
in the simulations shown, as the method quickly produces a strongly
centroidal mesh. We have explored this free parameter and found that
$2$ iterations is sufficient to produce quantitatively very similar
results.

Optionally, a limiter on the magnitude of the correction to the
velocity may be added to explicitly enforce the motion of the mesh
generating particles to be close to Lagrangian. We propose the following,
Galilean-invariant limiter, based on the local sound speed $c_{s,i}$
of the cell.
\begin{multline}
 \mathbf{\tilde{c}}_{i,J} \leftarrow (\mathbf{r}_i + \Delta t \mathbf{w}_{i,0}) +  
(\mathbf{\tilde{c}}_{i,J} - (\mathbf{r}_i + \Delta t \mathbf{w}_{i,0} )) \\
\times {\rm min}\left(
1,\frac{fc_{s,i}\Delta t }{\| \mathbf{\tilde{c}}_{i,J} - (\mathbf{r}_i + \Delta t \mathbf{w}_{i,0} ) \|}
  \right)
\end{multline}
That is, we restrict the motion of the cell to remain within a
circle of radius $fc_{s,i}\Delta t$, ($0<f<1$) around the first
predicted location of the mesh vertex (which is just predicted by the
fluid velocity).

The scheme may also be modified so that the centers-of-mass
$\mathbf{\tilde{c}}_{i,J}$ are calculated by weighting with the
density field, to bias the movement of the cells to regions of high
density. Alternatively, for a simpler implementation with the same
effect, the initial velocities in Equation~\ref{eqn:init} may be
biased by a density gradient term as in \cite{2012MNRAS.425.3024V}, to
move cells together towards regions of collapse.

An efficient implementation of the method does not require new
mesh reconstruction in each iteration. The center-of-mass of cell $i$
in the predictive step may be calculated by just using the points that
are neighbours of the cell at the beginning of the timestep (kicking them forward in time to build the predictive Voronoi cell). This strategy avoids additional in-circle tests. In cases
where the mesh topology changes, this is not strictly an exact
calculation of the center-of-mass, but the error is negligible since a
Voronoi diagram changes continuously.

\section{Numerical Tests}\label{sec:tests}

\subsection{Reduction of artificial secondary instabilities in the Kelvin Helmholtz instability}\label{sec:KH}

As a first test, we demonstrate how the mixing and formation of
secondary instabilities in the Kelvin Helmholtz instability (KHI)
changes visibly from the original KHI tests that were presented in the
{\sc Arepo} paper \citep{2010MNRAS.401..791S} using SSR. The setup
of the initial conditions of this shear flow is described in
\cite{2010MNRAS.401..791S}. The evolved KHI at $t=2.0$ is presented in
Fig.~\ref{fig:KH} with both the LR and SSR schemes, at low and high
resolutions. With the LR approach, we have enforced near-Lagrangian
motion with the correction limiter parameter of $f=0.1$. We ran the LR
simulation with the old MH integrator and the GG gradient
estimates. For comparison, we ran the SSR simulation in the MH
integrator and GG gradient estimates mode, as well as the new RK
integrator and LSF gradient estimates mode.

The SSR scheme shows evidence of mesh noise at both low and high
resolutions; there is structure present on the scale of the mesh size,
at either resolution. The LR method produces a result that is smooth
and in agreement with high resolution fixed-grid simulations in terms
of the secondary instabilities that develop (c.f. the static mesh runs
shown for comparison in Fig.~\ref{fig:KH}, or the static mesh runs in
Fig.~34 of \cite{2010MNRAS.401..791S}, or the static mesh
finite-element simulations in Fig.~8 of
\cite{2014MNRAS.437..397M}). Small scale secondary instabilities that
are present with the SSR scheme disappear with the new LR method. The
density field can be said to be well-resolved with the LR technique
because there are no structures on the length scale of the cell
size. It is worth pointing out that the solution is not exactly
symmetric at high resolution because truncation errors lead to chaotic
behaviour. We claim the secondary instabilities present in the old SSR
scheme runs are purely numerical artifacts, which we demonstrate below
with a second KHI test.

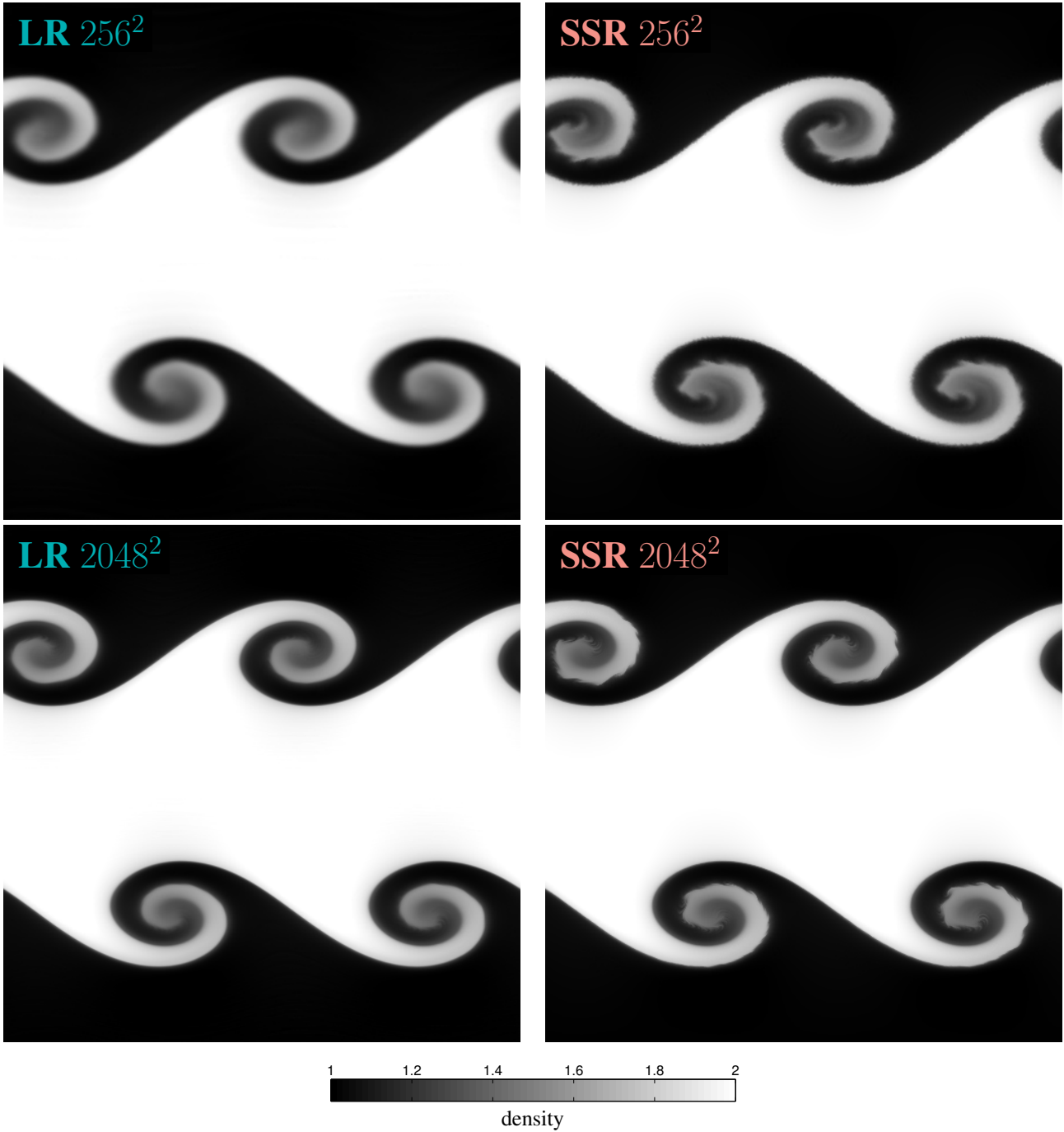
\begin{figure*}
\centering
\begin{tabular}{cc}
\begin{picture}(1,1)
\put(0,0) {\includegraphics[width=0.47\textwidth]{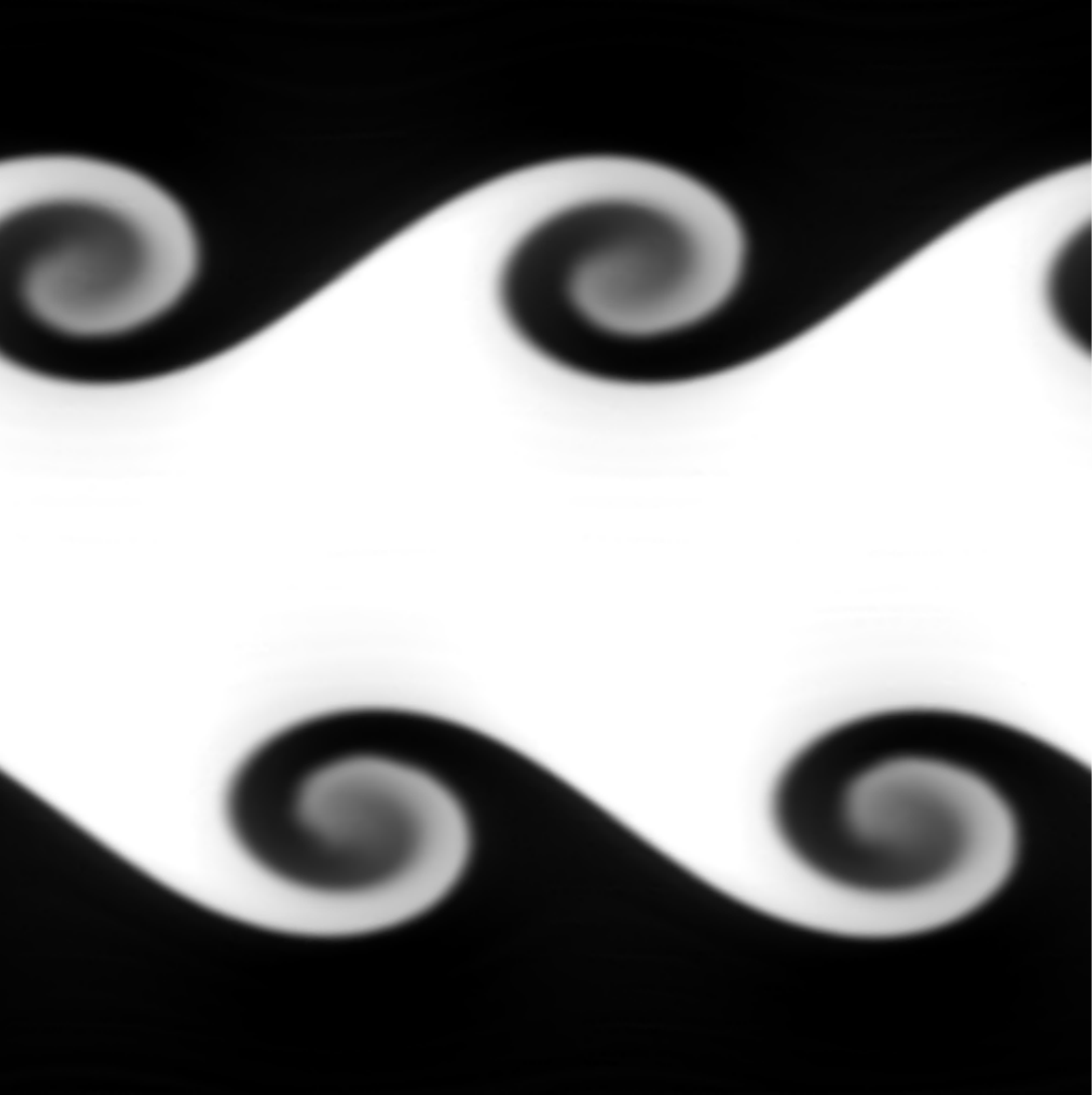} }
\put (0.015,0.92) {\colorbox{black}{\huge\bf\color{TealBlue} LR $256^2$}} 
\end{picture} &
\begin{picture}(1,1)
\put(0,0) {\includegraphics[width=0.47\textwidth]{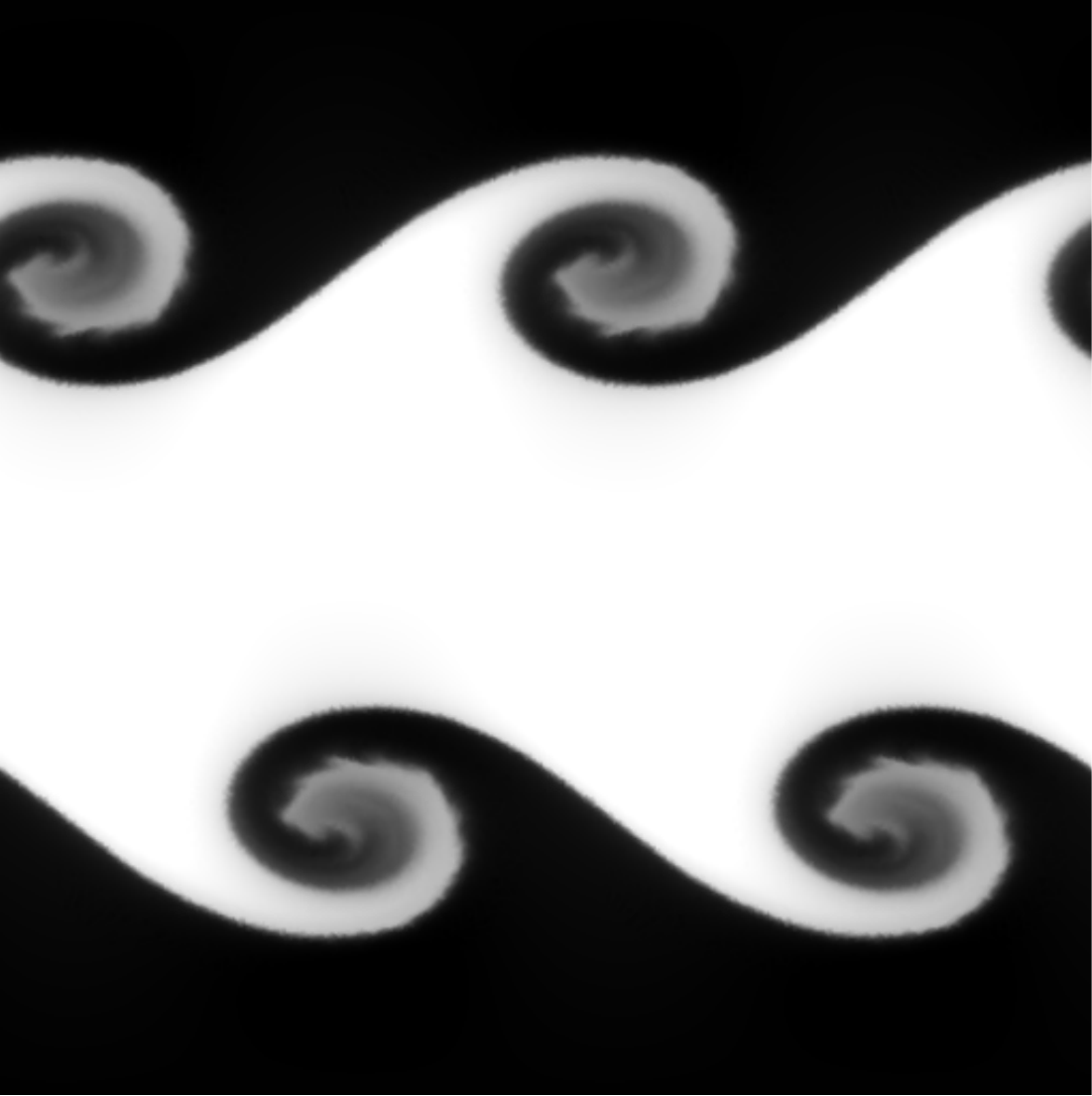} }
\put (0.015,0.92) {\colorbox{black}{\huge\bf\color{Salmon} SSR $256^2$}} 
\end{picture} \\
\begin{picture}(1,1)
\put(0,0) {\includegraphics[width=0.47\textwidth]{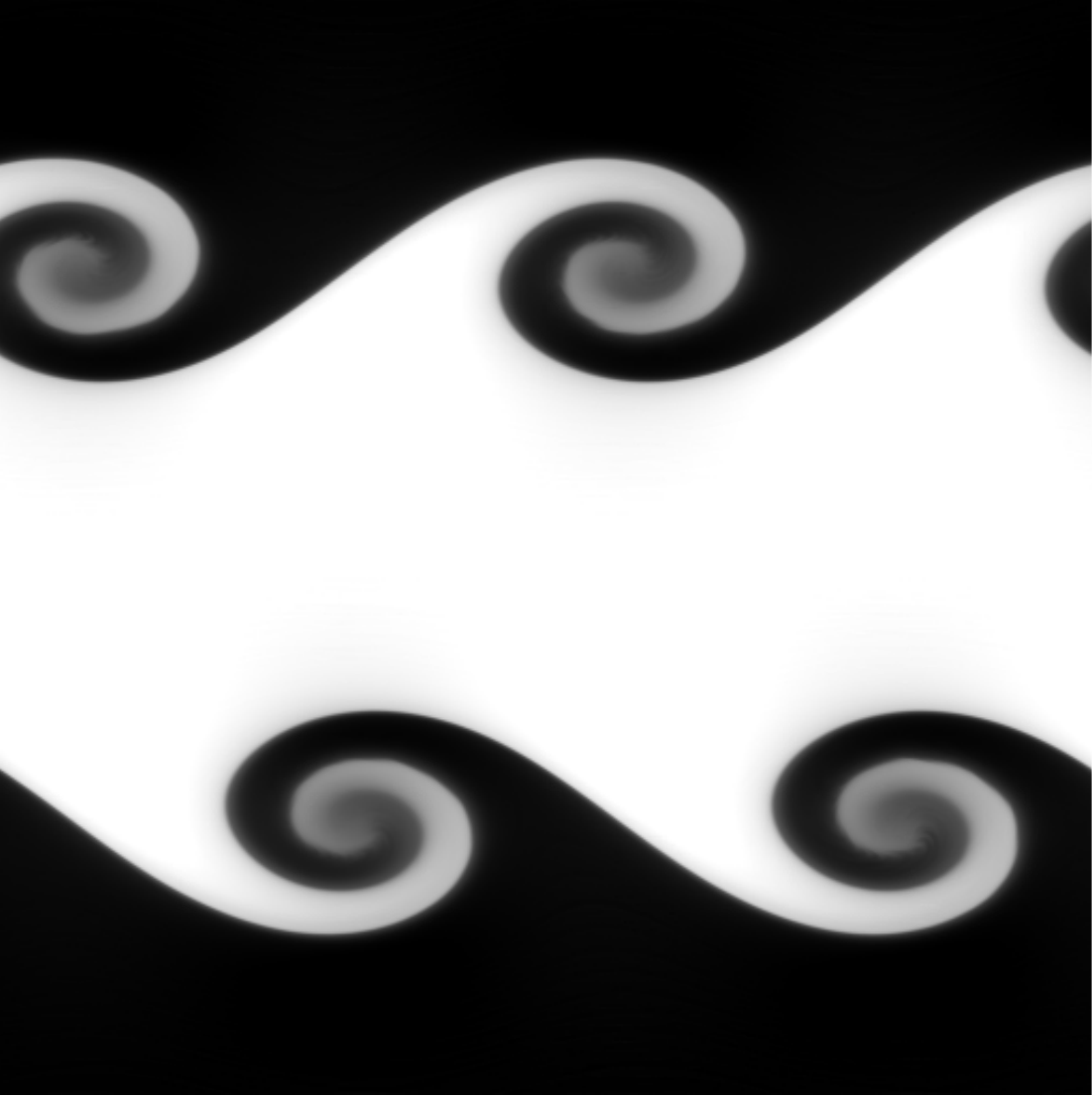} }
\put (0.015,0.92) {\colorbox{black}{\huge\bf\color{TealBlue} LR $2048^2$}} 
\end{picture} &
\begin{picture}(1,1)
\put(0,0) {\includegraphics[width=0.47\textwidth]{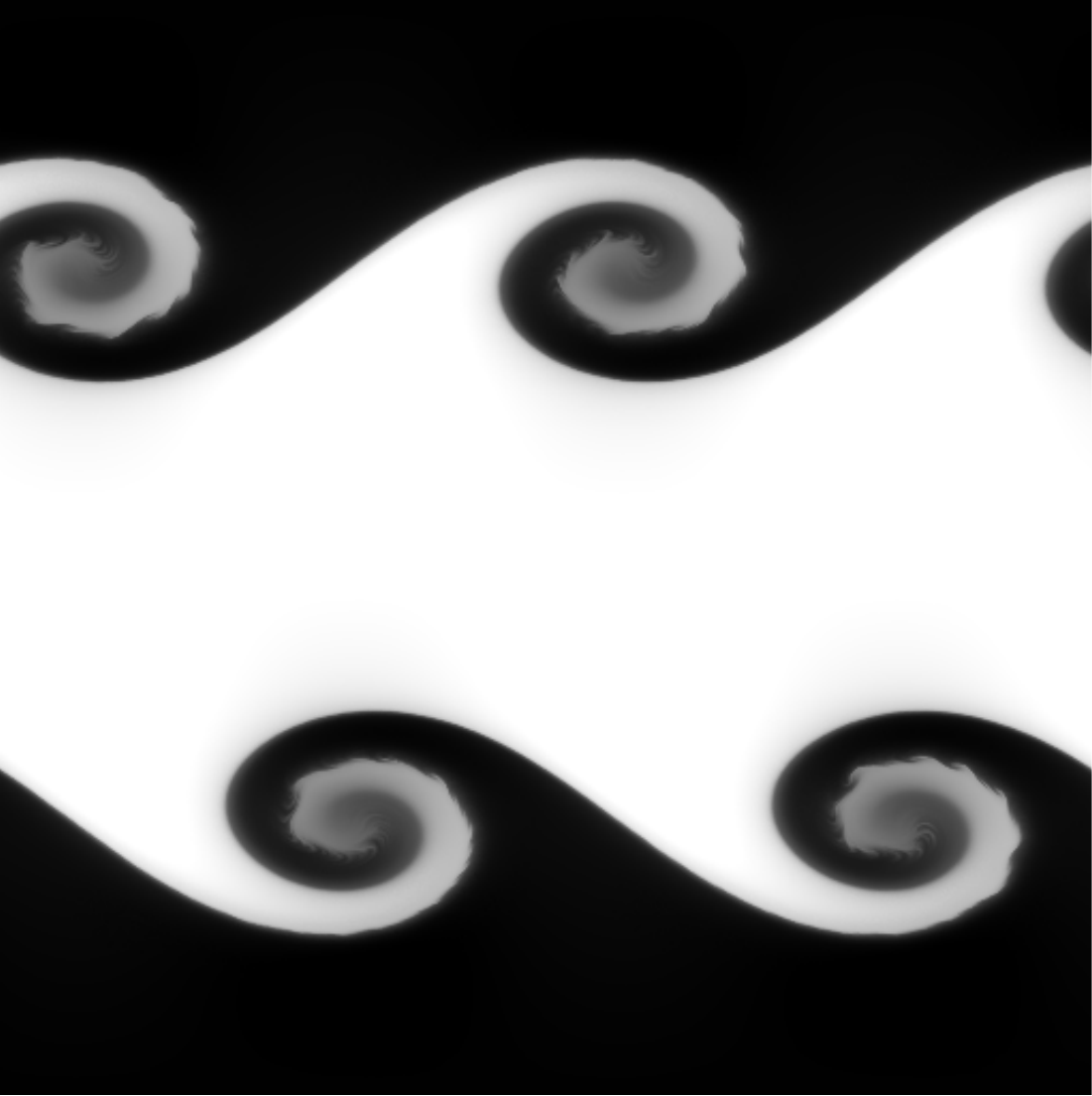} }
\put (0.015,0.92) {\colorbox{black}{\huge\bf\color{Salmon} SSR $2048^2$}}
\end{picture} \\
\multicolumn{2}{c}{\includegraphics[width=0.4\textwidth]{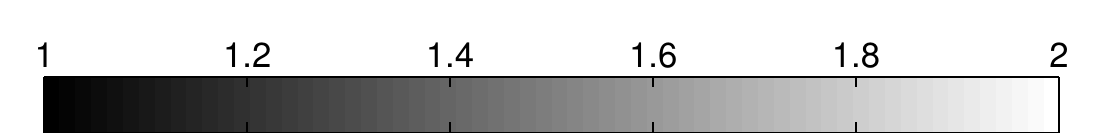}} \\
\multicolumn{2}{c}{{\large density}}
\end{tabular}
\caption{Comparison of KHI with the SSR and LR schemes at low and high resolution. The LR scheme eliminates secondary numerical instabilities.}
\label{fig:MN}
\end{figure*}

\begin{figure}
\centering
\includegraphics[width=0.47\textwidth]{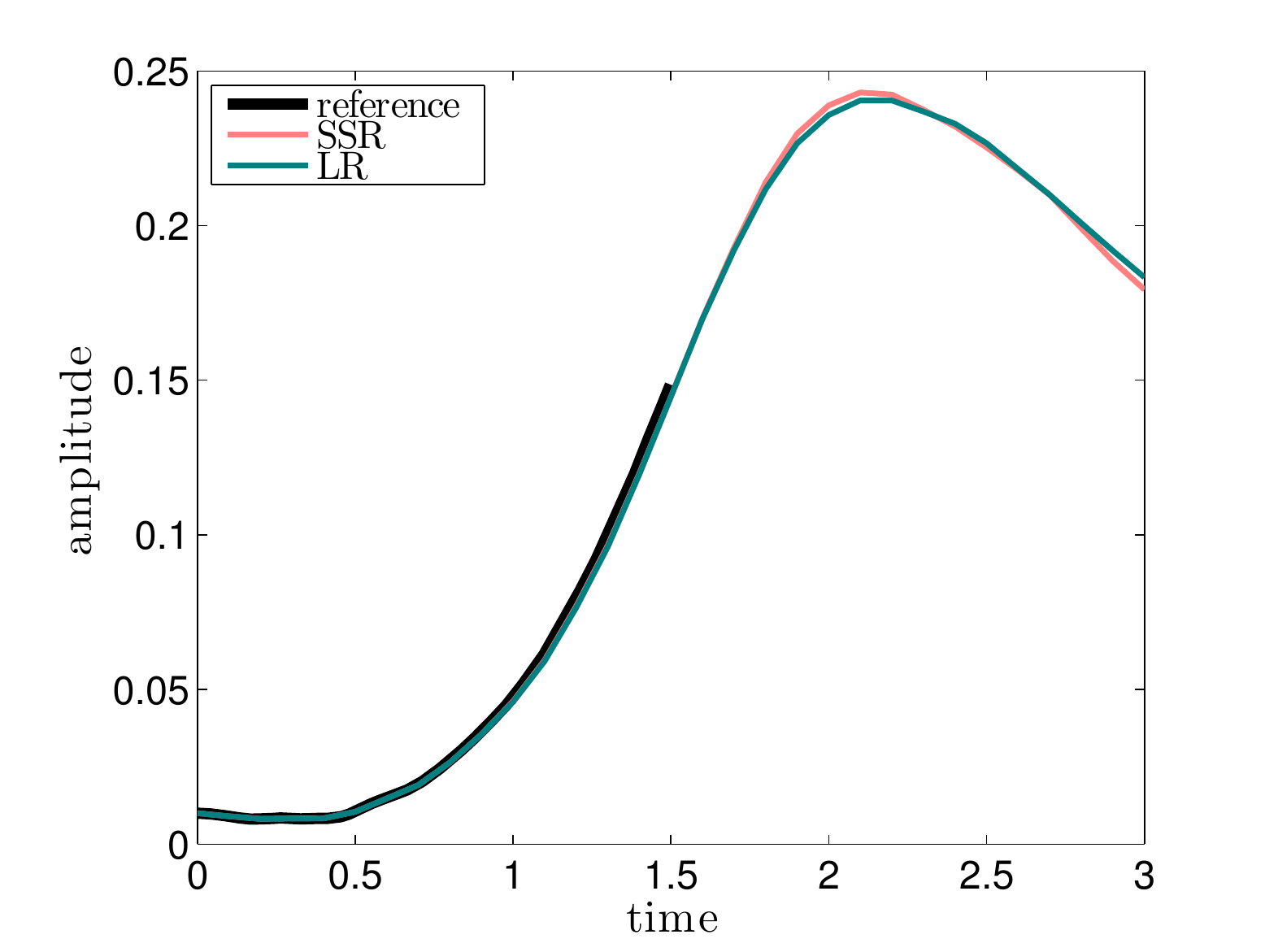}
\caption{Growth rates of the KHI simulated using the two regularization methods both show good agreement at early times with theory.}
\label{fig:MNgrowth}
\end{figure}

We investigate the KHI setup of \cite{2012ApJS..201...18M}, which is a
benchmark test in which only a single mode is excited and analytic
theory exists to predict the growth rate. Very high resolution
simulations of the instability are performed in
\cite{2012ApJS..201...18M} using the $6$-th order (in space) finite
difference {\sc Pencil} code. In Fig.~\ref{fig:MN}, we show that the
SSR scheme produces secondary instabilities on the scale of the cell
size, which are not present with the LR approach. The growth rate of the
primary mode of the instability of this test is still simulated
accurately with both regularization methods and agrees with theory, as
calculated and presented in Fig.~\ref{fig:MNgrowth}; only at late
times (in the non-linear regime) are small deviations observed in the
growth rate of the instability simulated with the two regularization
techniques, where we expect the LR scheme is providing more accurate
results due to the suppression of the secondary numerical
instabilities.

\begin{figure}
\centering
\includegraphics[width=0.47\textwidth]{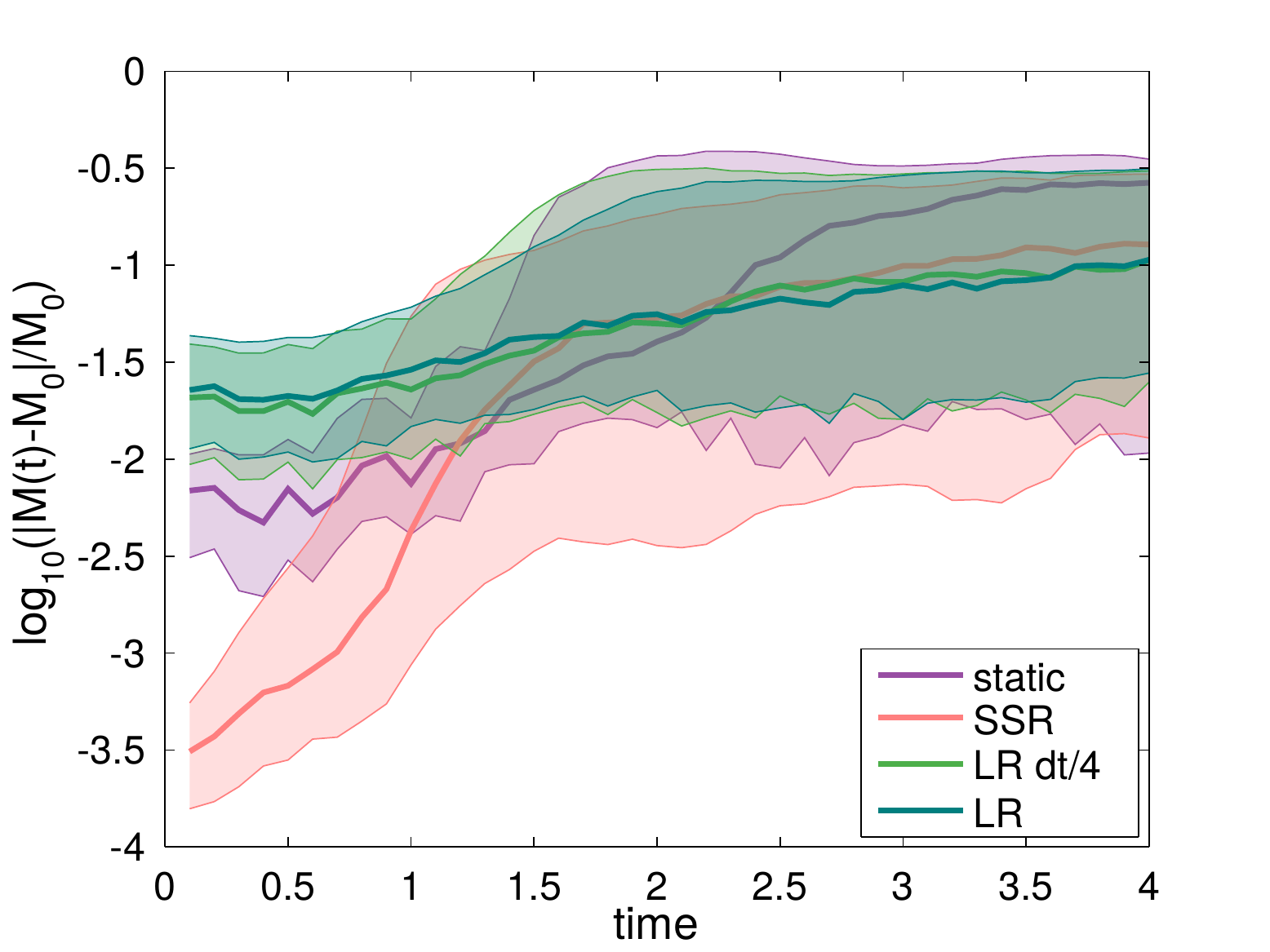}
\caption{Comparison of the mass of each cell as a function of time in a KHI simulation for the different regularization schemes (static mesh, SSR, LR with $1/4$ the time step, LR). The lines and shaded regions represent the quartiles of the distribution.}
\label{fig:KHlag}
\end{figure}

We explore the Lagrangian nature of the SSR and LR schemes in
Fig.~\ref{fig:KHlag}, where we plot the relative changes in the mass of
each cell as a function of time. In a purely Lagrangian scheme, there
is no mass exchange between particles. We see that the SSR and LR
schemes saturate to the same level of mass exchange, regardless of
whether the timestep is decreased by a factor of $4$. Hence, we
observe no strong implicit dependence of the deviation from Lagrangian
behaviour on the time step. We have also explored reducing the number of
Lloyd iterations from $5$ to $2$, and the mass exchange as function of
time and the density solution at $t=2$ do not change noticeably, so
they are not shown. The mass exchange on a static mesh saturates to a
higher level than the moving mesh approaches due to its Eulerian
nature.

\begin{figure*}
\centering
\begin{tabular}{cc}
\begin{picture}(1,1)
\put(0,0) {\includegraphics[width=0.47\textwidth]{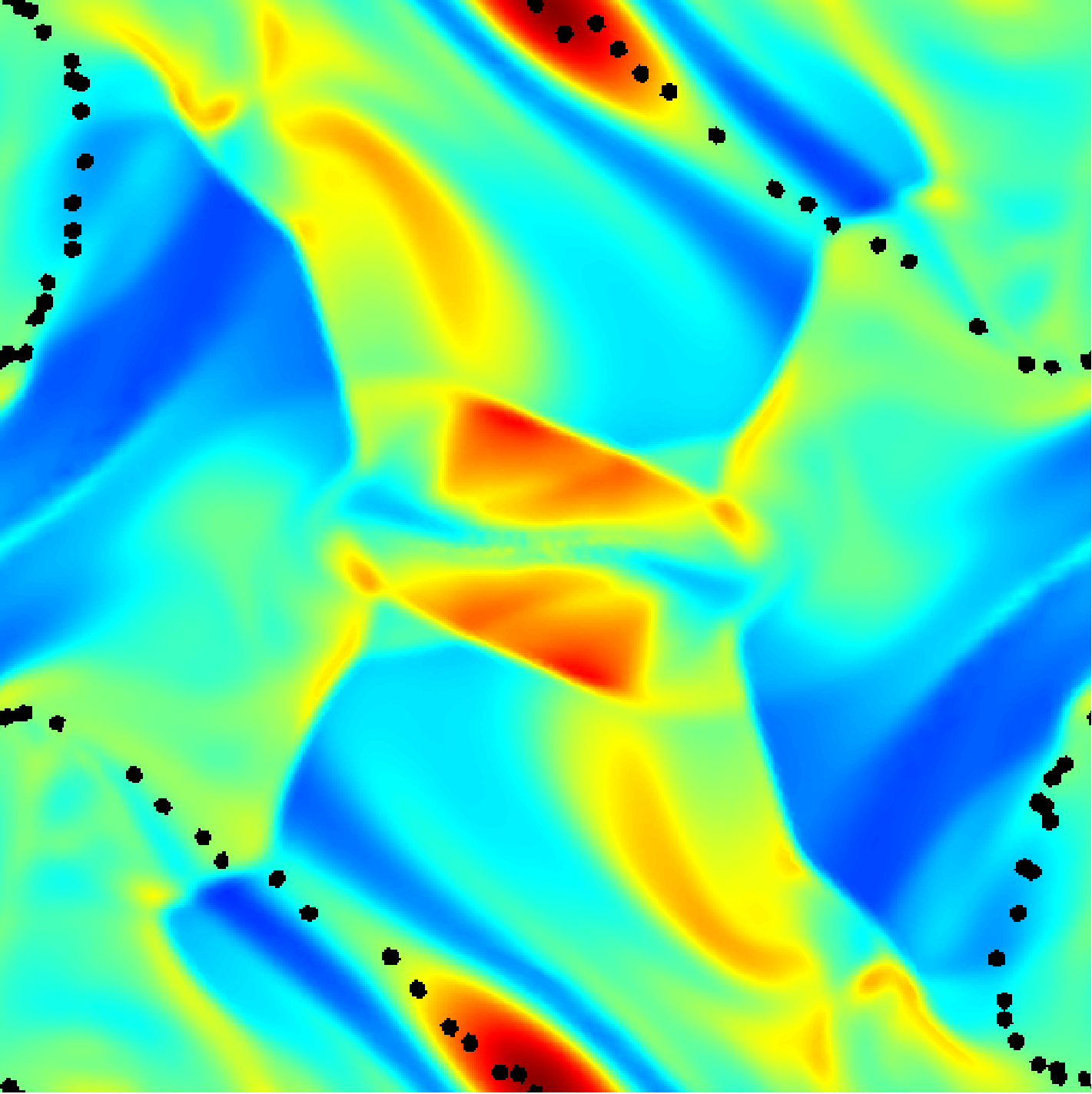} }
\put (0.015,0.92) {\colorbox{black}{\huge\bf\color{TealBlue} LR $128^2$}} 
\end{picture} &
\begin{picture}(1,1)
\put(0,0) {\includegraphics[width=0.47\textwidth]{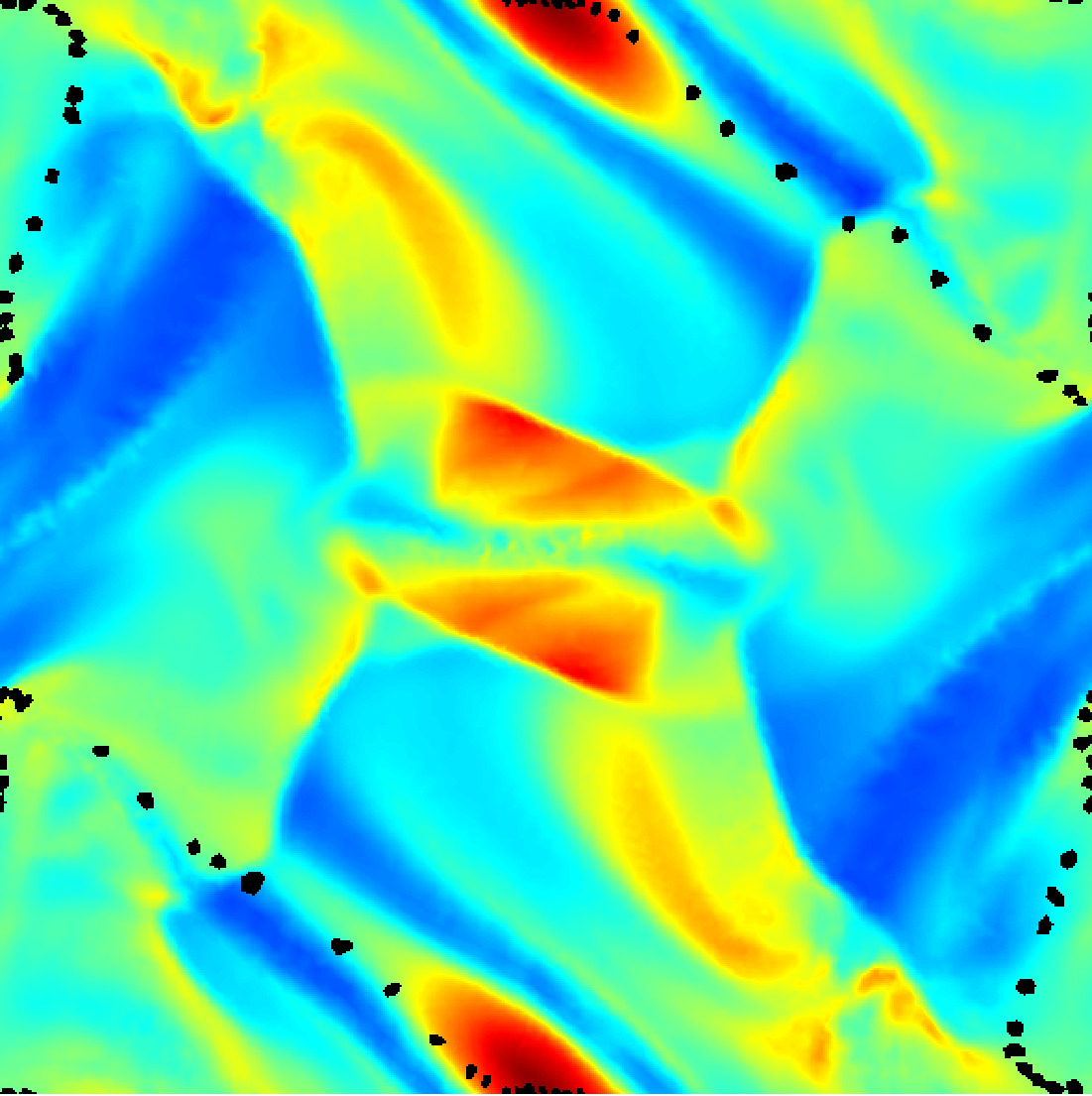} }
\put (0.015,0.92) {\colorbox{black}{\huge\bf\color{Salmon} SSR $128^2$}} 
\end{picture} \\
\multicolumn{2}{c}{\includegraphics[width=0.4\textwidth]{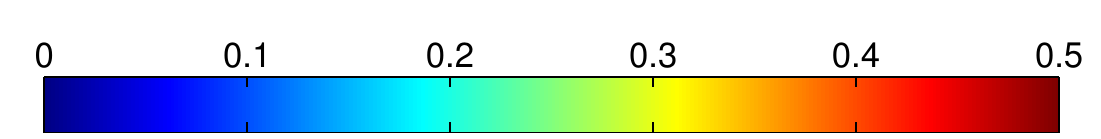}} \\
\multicolumn{2}{c}{{\large density}}
\end{tabular}
\caption{Density field at $t=0.5$ in the Orszag-Tang vortex simulated with the SSR and LR schemes. The LR scheme eliminates noise at locations of shocks and shear flow in the simulation. A number of cells are tagged in black (initially on the $y=0$ axis at $t=0$) to visually verify that they have travelled to similar locations at the end of the simulation.}
\label{fig:OT}
\end{figure*}

\subsection{Reduction of noise in the MHD Orszag-Tang Vortex}\label{sec:OT}

We next consider simulations of the Orszag-Tang vortex
\citep{1979JFM....90..129O}, a common MHD test problem that involves
supersonic shocks and decaying turbulence. The setup of the problem is
described in detail in \cite{2014MNRAS.442...43M}. We solve the system
using the constrained transport algorithm for a moving mesh described
in \cite{2014MNRAS.442...43M}. The outputs of the simulation at
$t=0.5$ are presented in Fig.~\ref{fig:OT}. The SSR scheme produces
numerical noise in the density field, particularly at the locations of
shocks and shear flows (compare the center and the
shocks in the four cardinal directions in the image). The degree of
these numerical noise errors can be changed by altering the mesh
regularization parameters of the SSR technique, but it is difficult a
priori to predict to what extent. For the tests presented here, we
chose the typical values suggested in \cite{2010MNRAS.401..791S}, and
also explored stronger regularization parameters, but found that the
noise does not go away completely because the corrections to the mesh
generating point locations are not applied in precisely the right
directions to keep the cells centroidal to high precision. These
numerical artifacts, however, are completely eliminated by the LR
scheme, which produces a clean, smooth solution. We note that there
are no parameters to adjust in the LR approach: the simulation
was run without the optional Lagrangian-enforcer (i.e., $f=\infty$).

In Fig.~\ref{fig:OT} we tagged cells initially on the $y=0$ axis at
$t=0$ to show where they finish at the end of the simulation, to
demonstrate that both regularization methods move the cells
approximately with the fluid flow, as they land approximately in the
same places, demonstrating that the scheme is quasi-Lagrangian. Thus, just
a tiny correction to the cell vertex velocities can have large
implications on the amount of grid noise in the solution: directed at
the right orientations they can be used to remove the grid noise
almost entirely.

\begin{figure}
\centering
\includegraphics[width=0.47\textwidth]{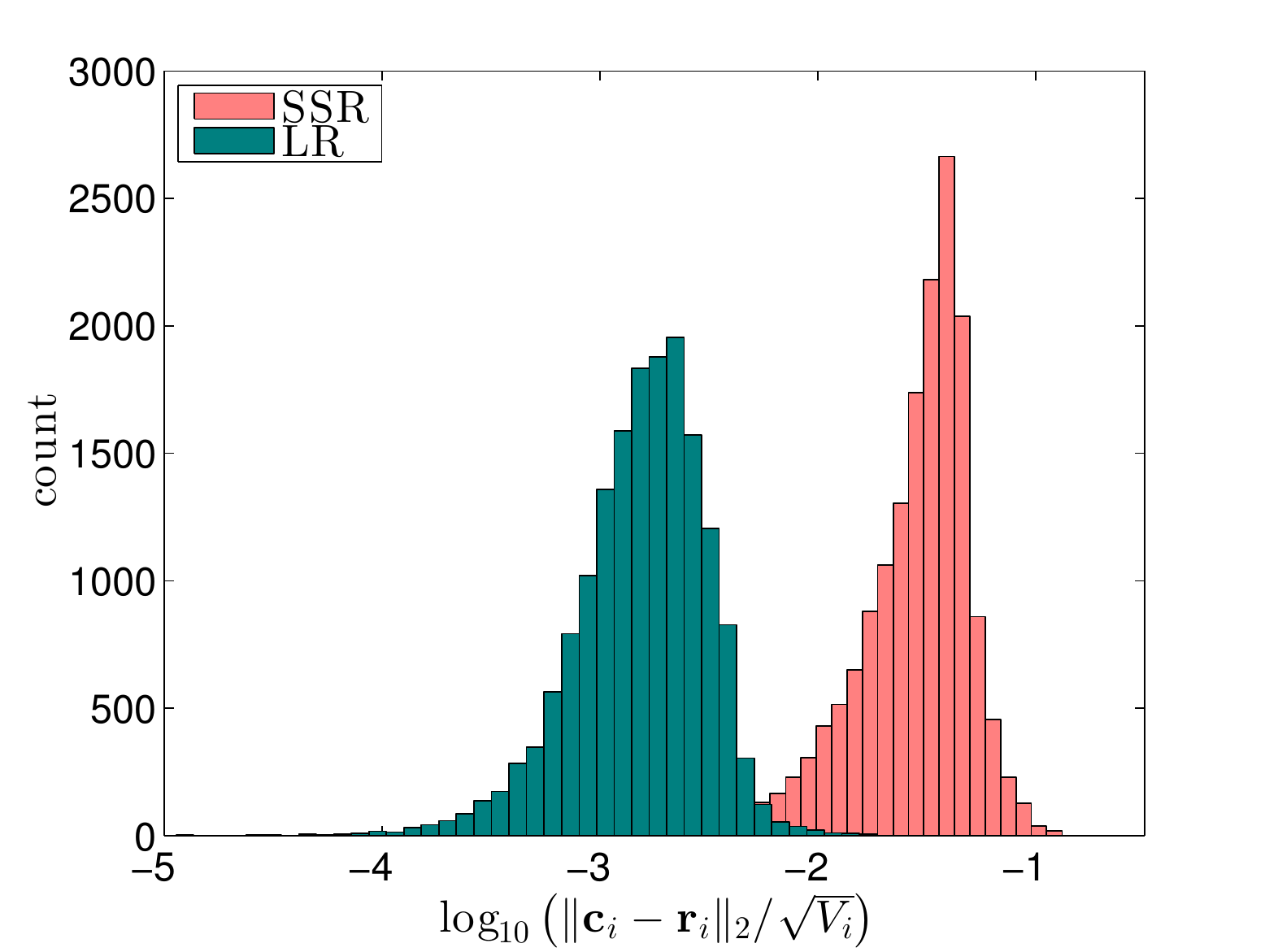}
\caption{A histogram of the relative center-of-mass, mesh generating point offset for the two regularization methods. The LR scheme shows over an order of magnitude improvement. The errors in the SSR technique can be quite large, $>10$ per cent.}
\label{fig:OThist}
\end{figure}

We verify that the LR scheme keeps the mesh regularized to a much
higher degree. A histogram of the relative center-of-mass, mesh
generating point offset for the two regularization techniques is shown in
Fig.~\ref{fig:OThist}. The LR method shows over an order of magnitude
improvement, and can be arbitrarily improved with more iterations in
the regularization algorithm. The offsets in the SSR scheme, on the
other hand, can be quite large ($>10$ per cent of the cell's effective
radius).

\subsection{Proper 2nd order convergence in Yee Vortex and Improved Angular Momentum Conservation}\label{sec:Yee}

With the LR scheme, we can achieve formal second order convergence with a moving Voronoi mesh code, which has been long-standing issue (see also \cite{2015arXiv150300562P} which achieves second order accuracy using an RK integrator and LSF gradient estimates). In many applications second-order convergence is not possible because the astrophysical flows generate shocks, and in the presence of these shocks the slope limiter reduces the order of accuracy to first-order in order to maintain numerical stability across discontinuities. However, second-order convergence should be expected for smooth flows. 

We achieve second order convergence by using the LR scheme and the time-symmetric RK integrator \citep{2015arXiv150300562P} but unlike \cite{2015arXiv150300562P} we can use GG gradient estimates since the mesh remains highly centroidal. The RK integrator averages a flux calculated with the mesh geometry at the beginning and end of the time step, instead of using just the mesh geometry at the beginning of the time step as the original MH integrator of {\sc Arepo} does. We demonstrate the ability of the code with the new LR scheme to show second-order convergence in the Yee vortex problem, a 2D isentropic, differentially-rotating, steady-state smooth flow \citep{1999JCoPh.150..199Y}.

The flow is described with parameters $T_{\rm inf} = 1$, $\beta=5$, $\gamma=1.4$, and box size $L=10$. The temperature, density, velocity, and pressure profiles are given as follows.
\begin{equation}
T(r) = T_{\rm inf} - \frac{(\gamma-1)\beta^2}{8\gamma \pi^2} {\rm e}^{1-r^2}
\end{equation}
\begin{equation}
\rho(r) = T^{1/(\gamma-1)}
\end{equation}
\begin{equation}
v_x(r) = -y \frac{\beta}{2\pi} {\rm e}^{(1-r^2)/2}
\end{equation}
\begin{equation}
v_y(r) = x\frac{\beta}{2\pi} {\rm e}^{(1-r^2)/2}
\end{equation}
\begin{equation}
P(r) = \rho^\gamma/\gamma
\end{equation}

\begin{figure}
\centering
\includegraphics[width=0.47\textwidth]{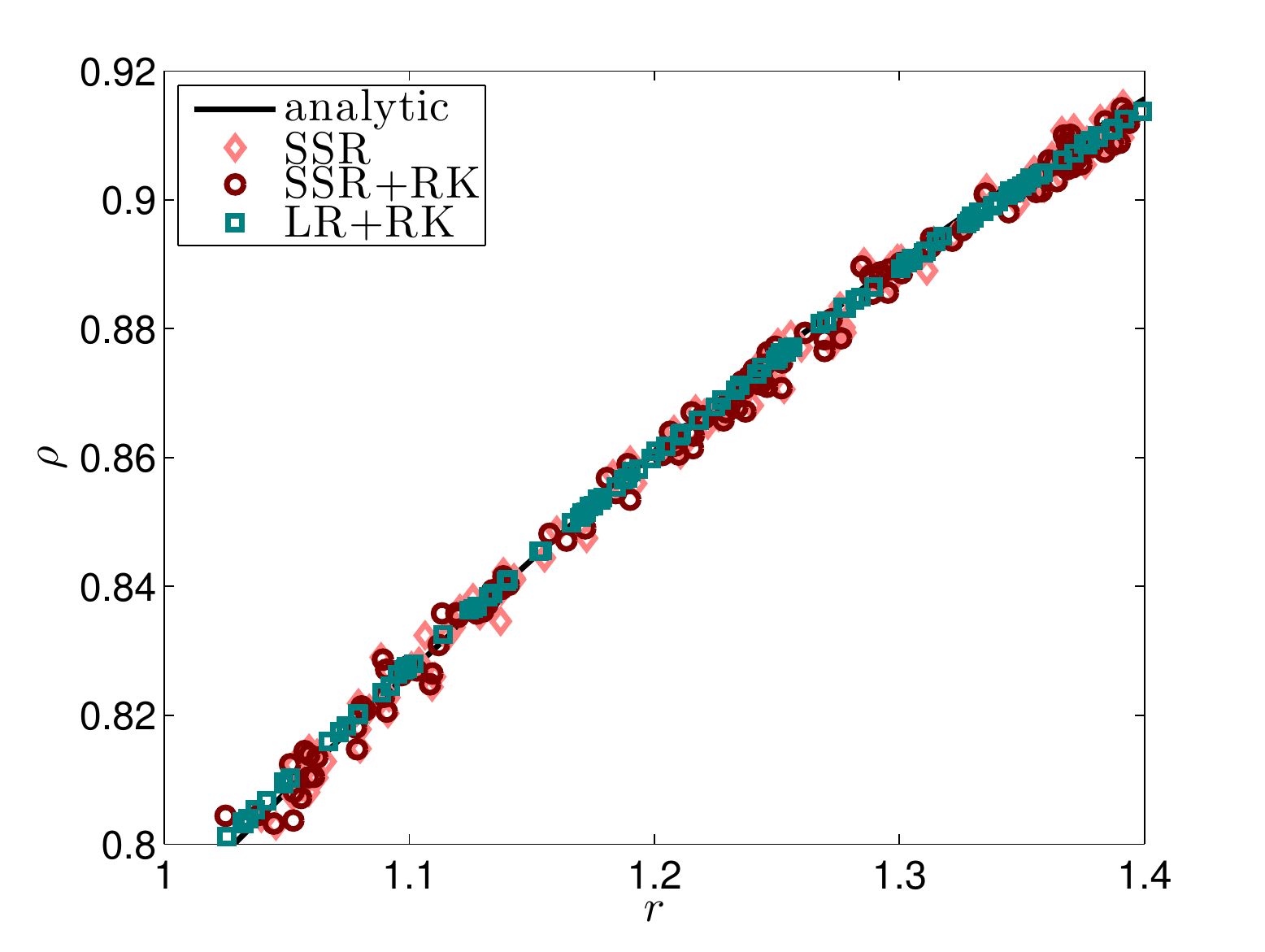}
\caption{The density profile of the Yee vortex at $t=10$, with the SSR and time stepping scheme, the RK time stepping scheme, and the RK scheme with the LR method. The LR eliminates the mesh noise.}
\label{fig:Yprofile}
\end{figure}

\begin{figure}
\centering
\includegraphics[width=0.47\textwidth]{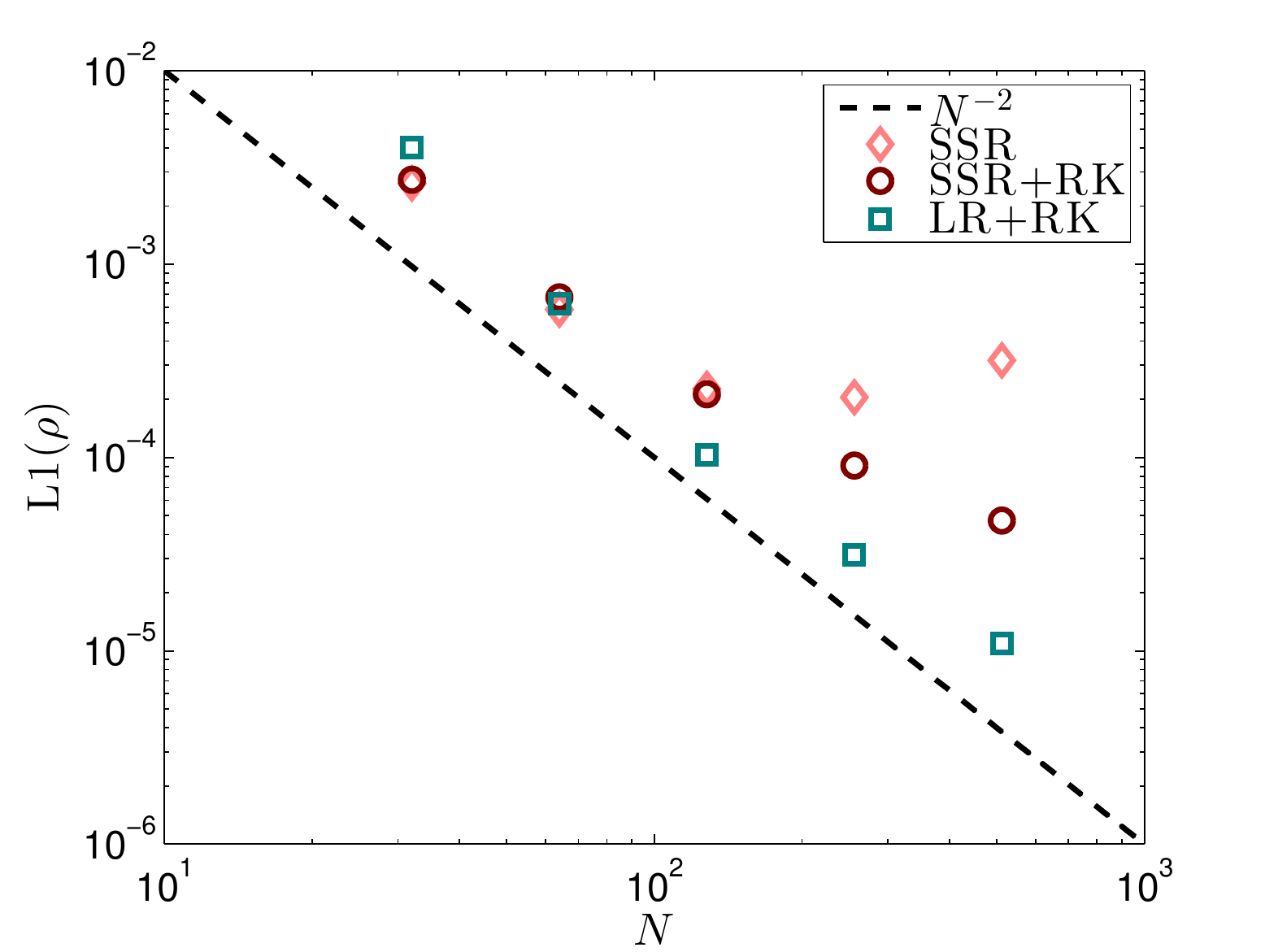}
\caption{Convergence analysis of the Yee vortex. Proper second-order convergence in the L1 norm is achieved with the LR and the RK time integrator.}
\label{fig:Yconv}
\end{figure}

We simulated this flow to $t=10$, a time at which the vortex has
differentially rotated several times. Fig.~\ref{fig:Yprofile} shows a
well behaved profile using the LR scheme, but a noisy one with the SSR
technique, even with the RK time integrator.  We see that due to the
presence of this mesh noise, the SSR method is unable to achieve
second-order convergence of the L1 norm error of the density profile,
as shown in the convergence plot Fig.~\ref{fig:Yconv}. Beyond some
fairly low resolution, the mesh noise dominates the error in this test
problem. However, this noise is removed with the LR scheme.

\begin{figure}
\centering
\includegraphics[width=0.47\textwidth]{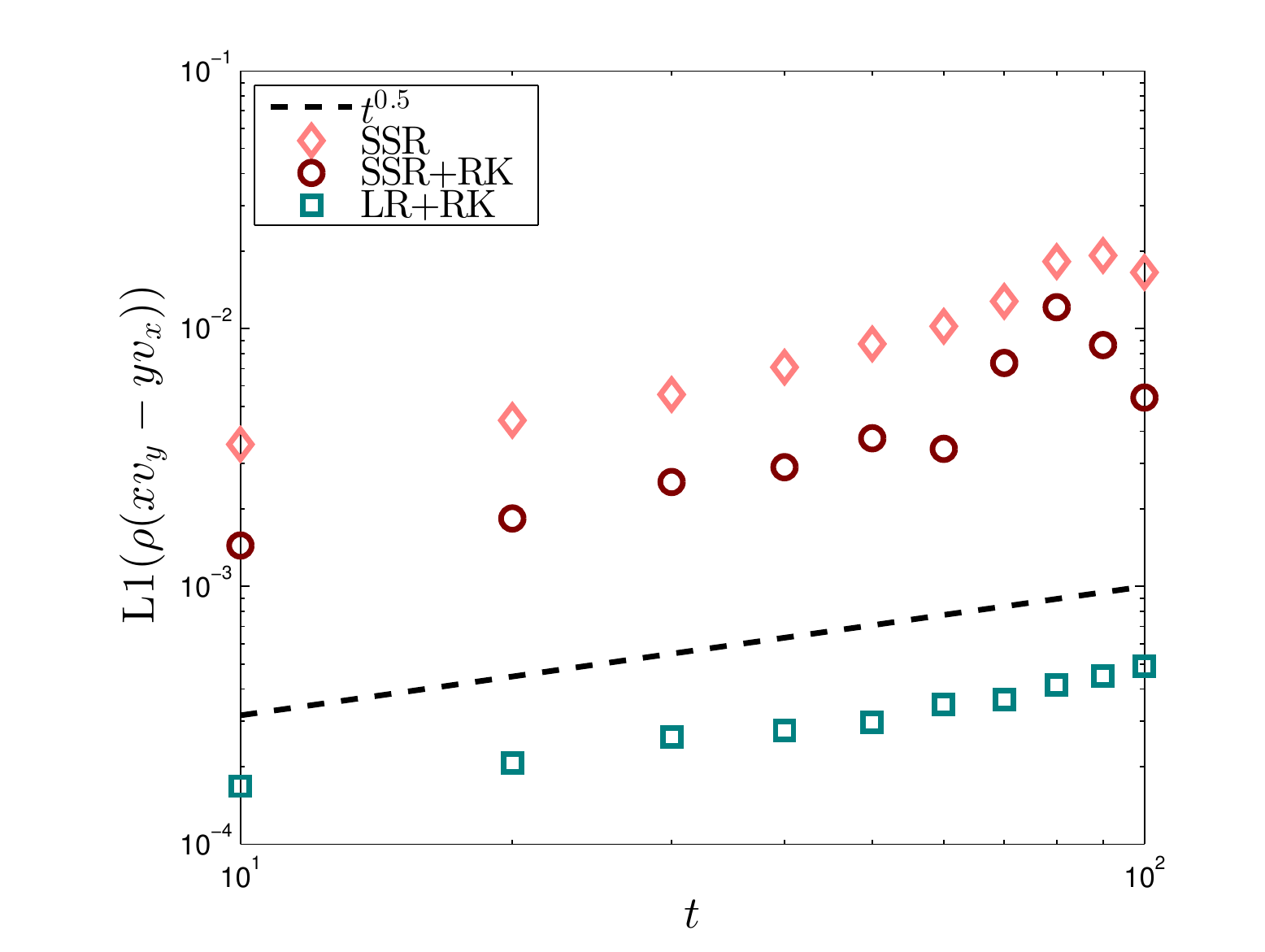}
\caption{The angular momentum profile error as a function of time. The error is reduced by an order of magnitude with the addition of the LR scheme, and grows in a predictable manner as $t^{0.5}$. }
\label{fig:YL}
\end{figure}

\begin{figure}
\centering
\includegraphics[width=0.47\textwidth]{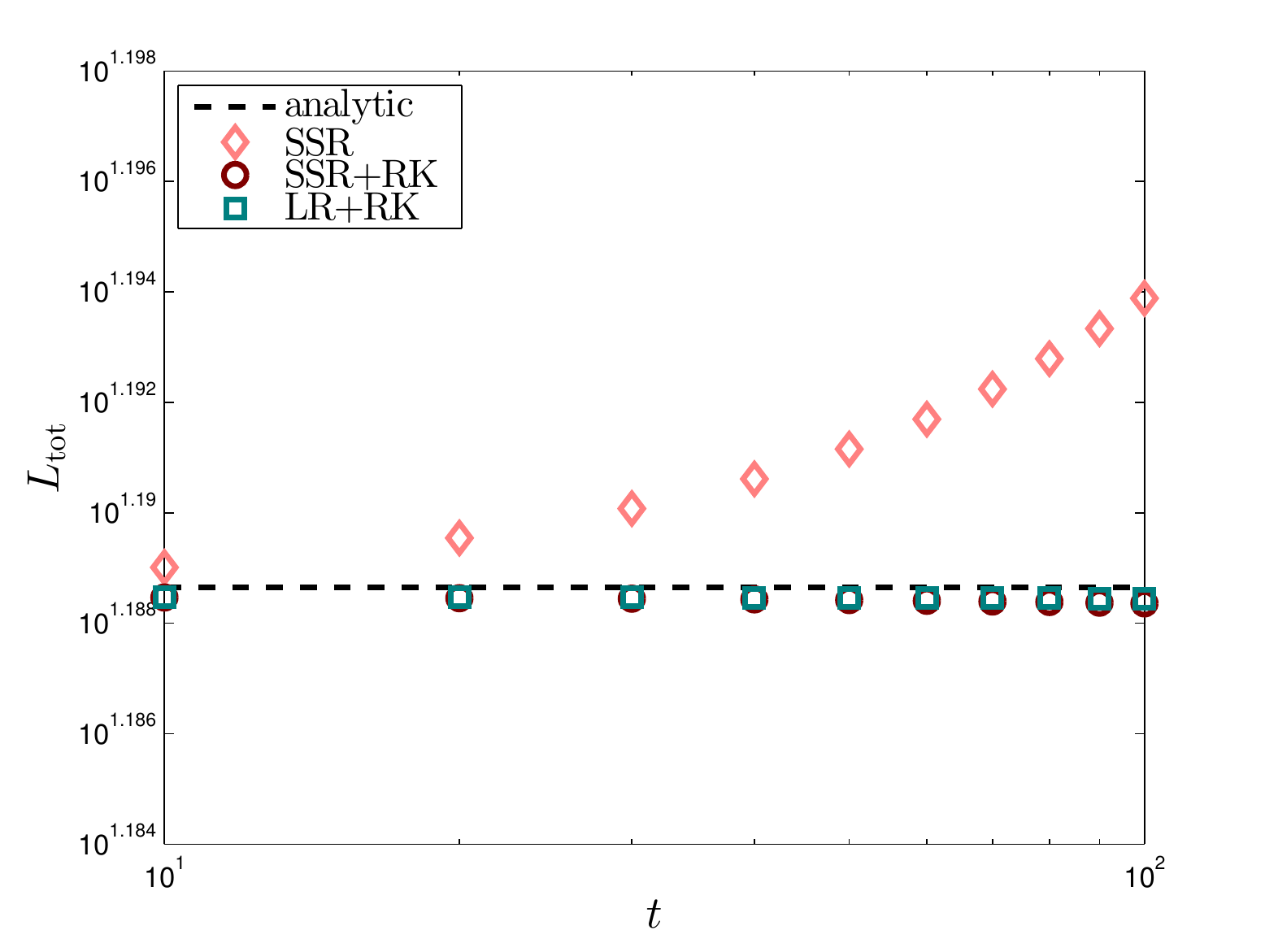}
\caption{The total angular momentum is well-behaved with the RK time integrator and the LR scheme, but may grow exponentially with the SSR scheme.}
\label{fig:YLtot}
\end{figure}

We simulate the flow much longer, up to $t=100$ to study the long term
behaviour of the angular momentum. The L1 norm error of the angular
momentum is reduced by an order of magnitude once the LR scheme is
added, as shown in Fig.~\ref{fig:YL}, and grows with time in a
predictable manner as $t^{0.5}$ for this problem. The angular momentum
without the LR scheme can fluctuate unpredictably due to mesh noise,
as also shown in Fig.~\ref{fig:YL}. The total angular momentum is well
behaved with the LR and RK time integrator improvements, as shown in
Fig.~\ref{fig:YLtot}, while with the original formulation of {\sc
  Arepo} it can grow non-linearly, which has previously placed some
limitations on the application of moving Voronoi mesh codes like {\sc
  Arepo} or {\sc TESS} for rotationally symmetric accretion
flows. With the LR scheme, we have improved the angular momentum
conservation of the moving mesh approach, and at the same time reduced
the mesh noise.

The well-behaved convergence with LR is achieved because there are no
errors in the gradient estimate present that are caused by deviations
from a fully centroidal mesh, which again result from the noise in the
movement of the mesh-generating points. We emphasise that the
combination of GG gradients and a centroidal mesh even guarantees
second order accurate gradients, whereas LSF gradients are only first
order accurate in general (for a Cartesian mesh, they are also second
order).

\subsection{Reduction of artificial power in turbulence test}\label{sec:turb}

\begin{figure}
\centering
\includegraphics[width=0.47\textwidth]{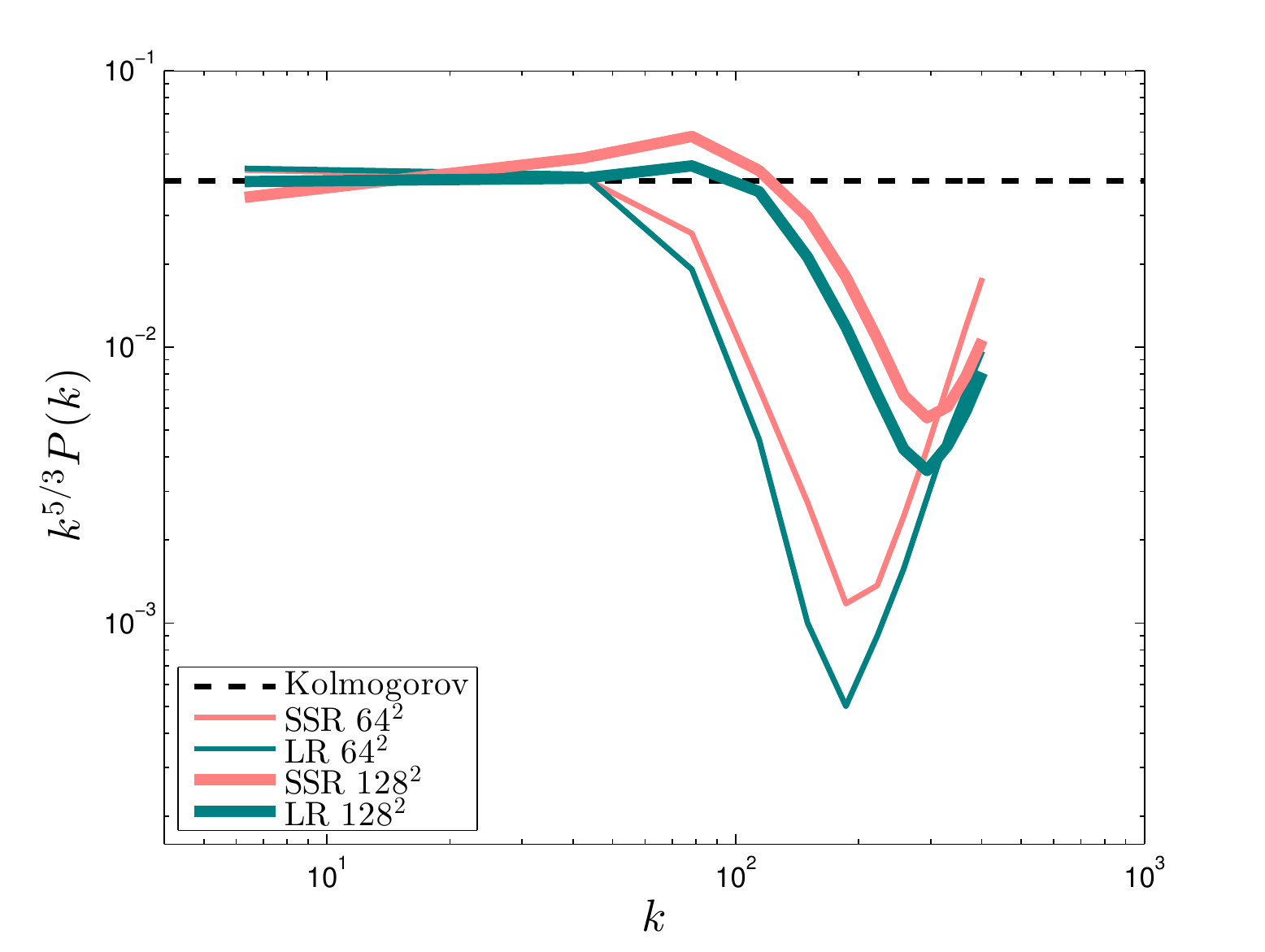}
\caption{Power-spectrum of driven subsonic turbulence for the two regularization methods. On the smallest spatial scales (large $k$), the SSR technique shows an excess build-up of power due to mesh noise, which is reduced with the LR scheme.}
\label{fig:Turb}
\end{figure}

We present the results of 3D subsonic driven turbulence, whose setup
is described in \cite{2012MNRAS.423.2558B}, and was also presented in
\cite{2014MNRAS.437..397M}, where the system was investigated using
finite-element methods. The energy power spectrum for such a turbulent
system is well described by a Kolmogorov power-law on the range of
spatial scales that are resolved by the mesh. However, the energy
cascade is artificially limited by the sizes of the cells, and there
is excess artificial accumulation of energy on these scales. This
excess build-up typically has tended to be larger in moving mesh codes
than in static mesh codes, with mesh noise being a likely culprit as
it adds fluctuations on the length scale of the cell size.  In
Fig.~\ref{fig:Turb}, we show that the artificial accumulation of
energy on the scale of the cells is reduced with the LR approach.  The
LR run used a Lagrangian enforcing parameter of $f=0.1$, same as the
KHI simulations, chosen so that the relative offsets between
centres-of-mass and mesh generating points are reduced by at least an
order of magnitude from the SSR scheme.

\begin{figure}
\centering
\includegraphics[width=0.47\textwidth]{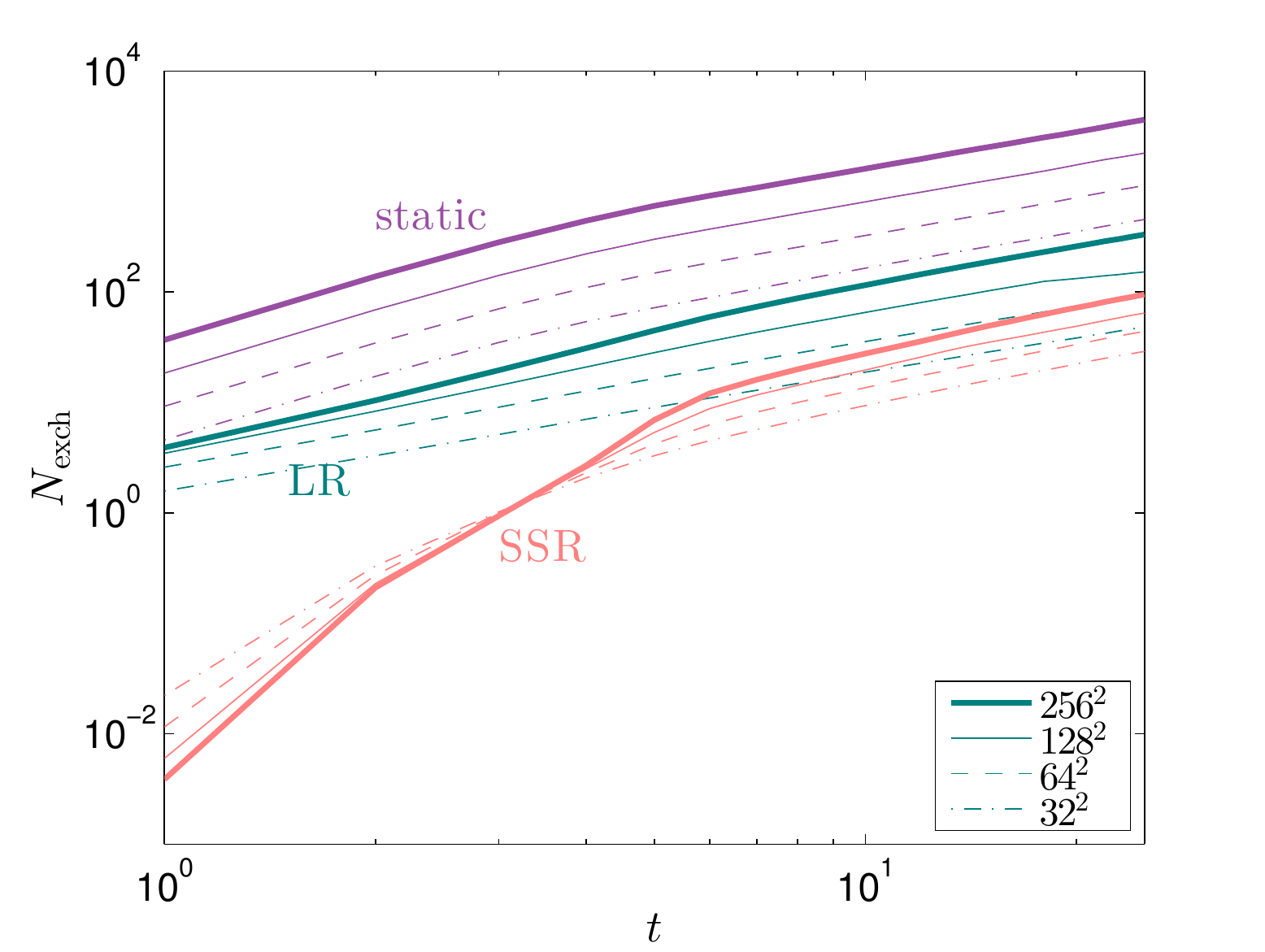}
\caption{The number of exchanges between cells, $N_{\rm exch}$, that
Monte Carlo tracers experience as a function of time in the various
turbulent box runs using different cell regularization schemes and at different resolutions. 
A lower number of exchanges indicates a simulation that is closer to fully-Lagrangian. Using a moving mesh instead of a static mesh reduces the 
number of exchanges by over an order of magnitude. The LR method, in keeping the mesh centroidal to a high degree, is somewhat less Lagrangian than the SSR scheme.}
\label{fig:TurbExch}
\end{figure}

We study the Lagrangian nature of the simulations by adding passive
Monte Carlo tracers in the turbulence tests, as in
\cite{2013MNRAS.435.1426G}. The number of exchanges between cells,
$N_{\rm exch}$, that Monte Carlo tracers experience as a function of
time, should be ideally zero in a fully-Lagrangian code. We plot the
evolution of this quantity in Fig.~\ref{fig:TurbExch}. The LR and SSR
methods show greatly reduced exchanges from a static mesh run, as
expected. For a Lagrangian enforcing parameter of $f=0.1$, the LR
scheme exhibits somewhat more exchanges than the SSR scheme. The exchange
number can be reduced by decreasing $f$ to make the Lagrangian
enforcement stronger, at the cost of weakening the strongly centroidal
condition.

\subsection{Adiabatic jet simulations with KHI}\label{sec:jet}

\begin{figure}
\centering
\begin{picture}(1,.47)
\put(0,0) {\includegraphics[width=0.47\textwidth]{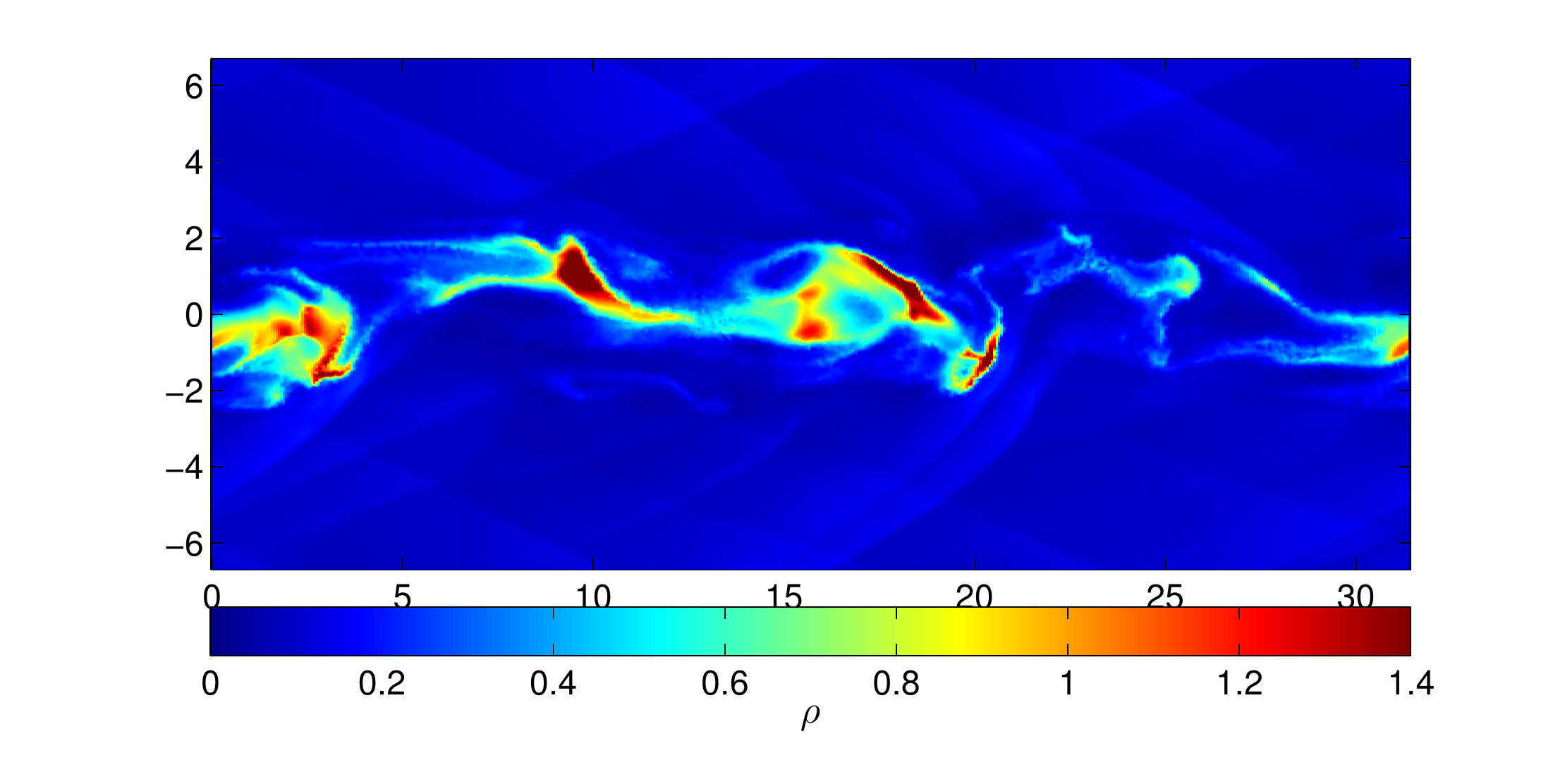} }
\put (0.15,0.39) {\colorbox{black}{\huge\bf\color{Salmon} SSR}} 
\end{picture} 
\begin{picture}(1,.47)
\put(0,0) {\includegraphics[width=0.47\textwidth]{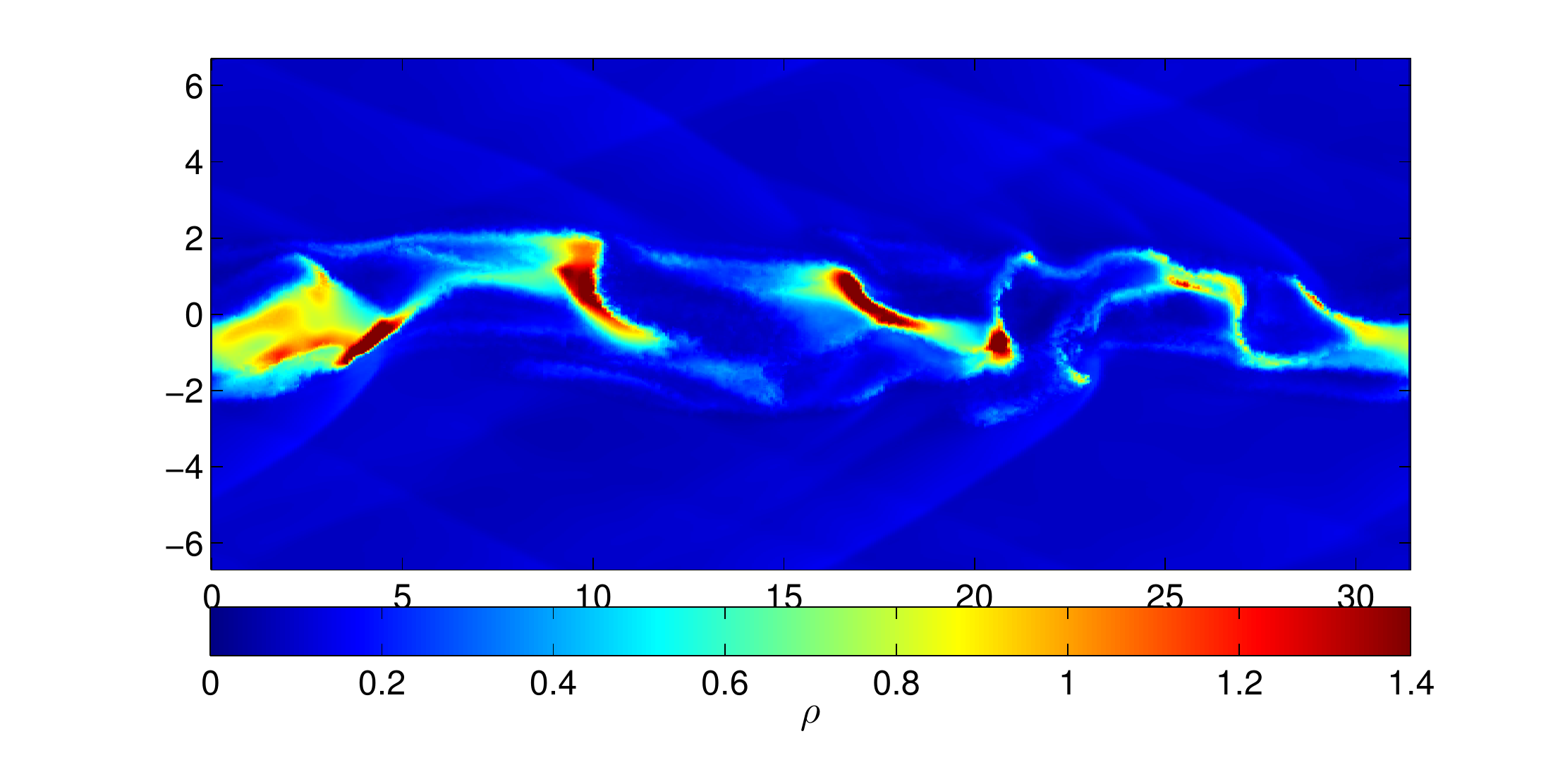} }
\put (0.15,0.39) {\colorbox{black}{\huge\bf\color{TealBlue} LR}} 
\end{picture} 
\includegraphics[width=0.47\textwidth]{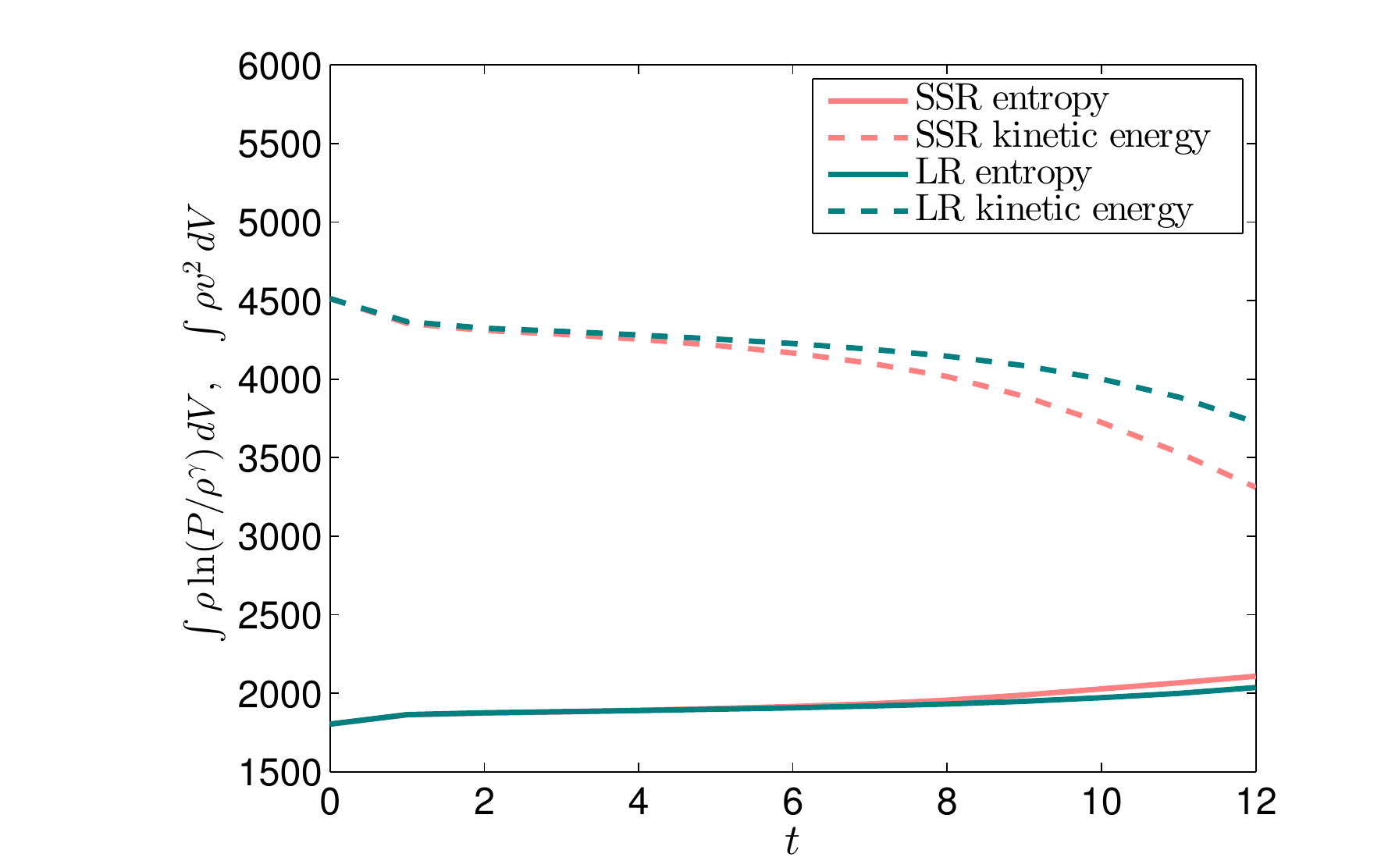}
\caption{Simulation of an $M=10$, heavy, adiabatic jet with the two regularization methods. The cross-sections are at time $t=7$ sound-crossing times. Entropy generation is higher at late times with the SSR scheme than with the LR scheme.}
\label{fig:jet}
\end{figure}

As a test of the new regularization method in an astrophysical context
where it is expected to have an impact, we present simulations of
heavy, supersonic, adiabatic jets that have a helical mode excited
which triggers the KHI, following the setup of
\cite{1998A&A...333.1117B,2000A&A...360..795M}. This situation is
ideal for the study of different phases of the temporal evolution as a
function of sound-crossing time. The jet has a surrounding gas density
to jet density ratio of $\nu=0.1$, and a supersonic Mach number of
$M=10$. Our simulations are in 3D and use $5$ million cells. 
In Figure~\ref{fig:jet}, we
present the results of the simulation at a point where the KHI has
developed and shredded the jet.  The LR scheme
shows lower level of mesh noise and entropy generation than the SSR
technique at late times in the simulation.

\subsection{Isolated galaxy formation}\label{sec:disc}

\begin{figure*}
\centering
\begin{picture}(0.6,0.6)
\put(0,0) {\includegraphics[width=0.3\textwidth]{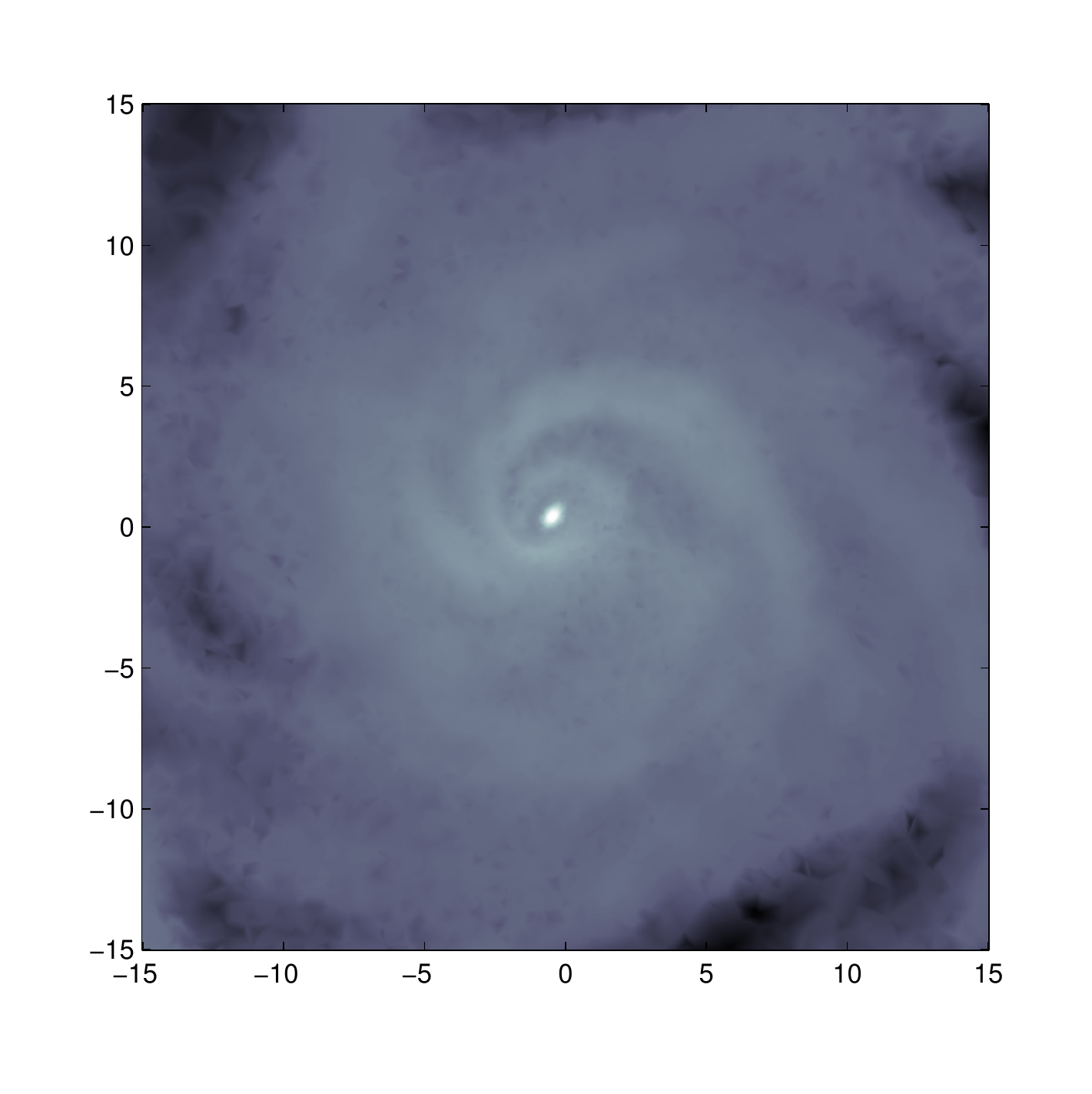} }
\put (0.1,0.52) {\colorbox{black}{\Large\bf\color{Salmon} SSR}}
\end{picture} 
\begin{picture}(0.6,0.6)
\put(0,0) {\includegraphics[width=0.3\textwidth]{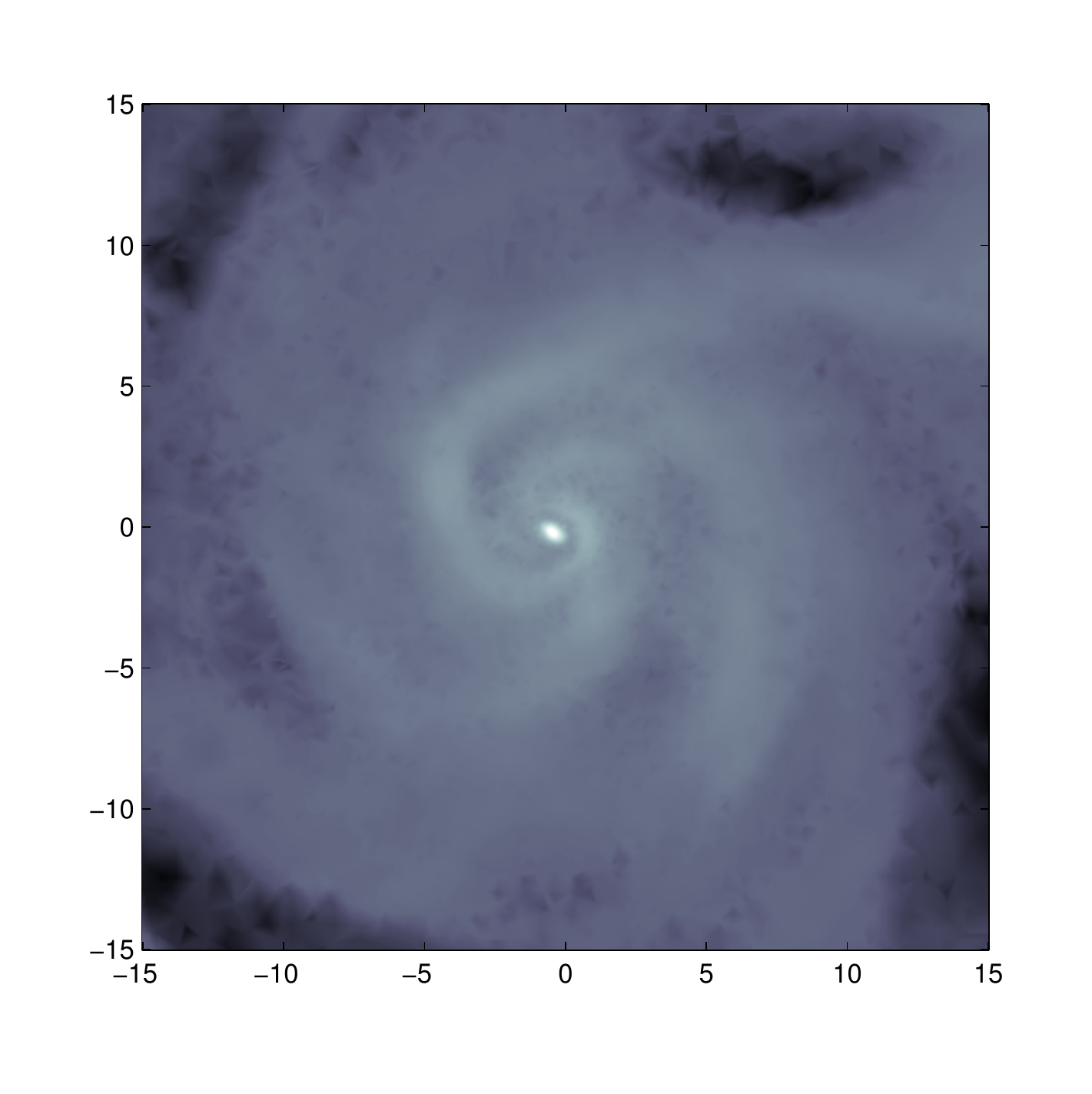} }
\put (0.1,0.52) {\colorbox{black}{\Large\bf\color{TealBlue} OR}}
\end{picture} 
\includegraphics[width=0.39\textwidth]{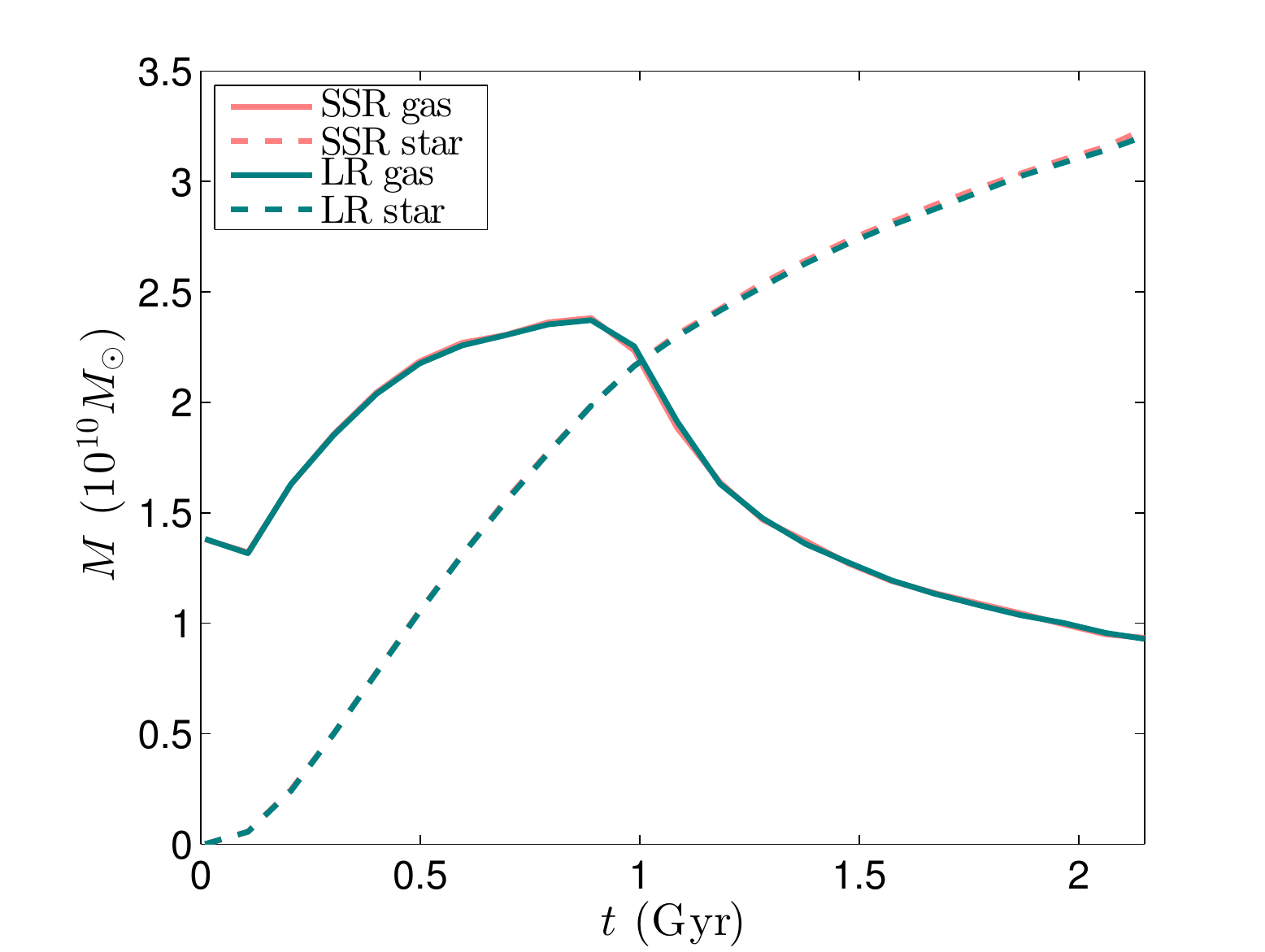}
\caption{Simulation of the formation of an isolated magnetic disc using the two regularization schemes. The first two panels show a slice of the gas density field, while the third panel shows the time evolution of mass of stars and gas within a $15$~kpc radius. Regularization does not affect the global properties of the simulation. The density scale covers $-5\leq \log_{10}\rho \leq-0.8$.}
\label{fig:disc}
\end{figure*}

Finally, we simulate the formation of an isolated disk galaxy with magnetic
fields similar to the Milky Way, following the setup of
\cite{2013MNRAS.432..176P}. The initial gas sphere in the simulation
is in hydrostatic equilibrium (without radiation), but collapses due
to radiative cooling. The gas has some initial angular momentum, and
settles into a dense, rotationally supported disk. Regions in the disc
can fragment and are allowed to form stars. In this simulation with
collapse, we bias the initial velocities of the mesh generating points
by the density gradient term as in \cite{2012MNRAS.425.3024V}, before
applying the Lloyd regularization, in order to closely follow the
collapse. We present the results of the simulations under the LR and
SSR schemes as well as time evolution of some global properties in
Figure~\ref{fig:disc}, and find the results are unaffected by the
regularization because mesh noise errors are sub-dominant. The
simulations have a mass resolution of $10^6~M_\odot$.

\section{Concluding remarks}\label{sec:conclusion}

In summary, we have presented a new regularization scheme, Lloyd
Regularization (LR) for moving Voronoi mesh codes that significantly
reduces mesh noise and improves convergence and angular momentum
conservation. These advantages are gained because the new LR scheme
keeps the cells centroidal to a high degree by applying a
forward-predicting Lloyd iteration correction to the mesh generating
point. The regularization scheme will improve the science applications
of moving mesh codes used in astrophysics, particularly in systems
where shear flows exist. The LR scheme allows moving mesh codes to
capture the physics of fluid instabilities more accurately than
earlier methods, which are thought to be important for jet dynamics
and gas stripping in galaxies.  The correct level of turbulence and
entropy generation is also improved with the reduction of mesh noise
through this method.  The improvements are most relevant for shear
flows, and complements the recent moving-mesh improvements for smooth
solutions of \cite{2015arXiv150300562P}.

\section*{Acknowledgements}
This material is based upon work supported by the National Science
Foundation Graduate Research Fellowship under grant no. DGE-1144152
(PM). LH acknowledges support from NASA grant NNX12AC67G and NSF award
AST-1312095. SG acknowledges support for program number
HST-HF2-51341.001-A provided by NASA through a Hubble Fellowship grant
from the STScI, which is operated by the Association of Universities
for Research in Astronomy, Incorporated, under NASA contract
NAS5-26555. RP acknowledges support by the European Research Council
under ERC-StG grant EXAGAL-308037 and by the Klaus Tschira Foundation.

\bibliography{mybib}{}

\bsp
\label{lastpage}
\end{document}